\documentclass[prl,twocolumn,showpacs,superscriptaddress,amssymb]{revtex4}

\usepackage{psfrag,graphicx}
\usepackage{dcolumn}
\usepackage{bm}
\usepackage{amsfonts,amssymb,amsmath}        % for math symbols.
\usepackage{color}
\usepackage{xcolor}
\usepackage{slashed}

\newcommand{\be}{\begin{equation}}
\newcommand{\ee}{\end{equation}}
\newcommand{\bq}{\begin{eqnarray}}
\newcommand{\eq}{\end{eqnarray}}

\newcommand{\pp}[2]{\frac{\partial#1}{\partial #2}}
\newcommand{\cev}[1]{\reflectbox{\ensuremath{\vec{\reflectbox{\ensuremath{#1}}}}}}
\newcommand{\eqtext}[1]{\ensuremath{\stackrel{\text{#1}}{=}}}

\usepackage{tikz}
\usetikzlibrary[shapes.arrows]
\usetikzlibrary{shapes}
\usetikzlibrary{arrows, decorations.markings}
\tikzstyle{vecArrow} = [thick, decoration={markings,mark=at position
   1 with {\arrow[semithick]{open triangle 60}}},
   double distance=1.4pt, shorten >= 5.5pt,
   preaction = {decorate},
   postaction = {draw,line width=1.4pt, white,shorten >= 4.5pt}]
\tikzstyle{innerWhite} = [semithick, white,line width=1.4pt, shorten >= 4.5pt]

\usepgflibrary{arrows}
\usetikzlibrary{arrows} 
\usetikzlibrary{positioning}
\usetikzlibrary{calc}
\usetikzlibrary{decorations.pathmorphing}
\usetikzlibrary{backgrounds}

\makeatletter
\pgfdeclareradialshading[tikz@ball]{ball}{\pgfqpoint{5bp}{10bp}}{%
 color(0bp)=(tikz@ball!30!white);
 color(9bp)=(tikz@ball!75!white);
 color(18bp)=(tikz@ball!90!black);
 color(25bp)=(tikz@ball!70!black);
 color(50bp)=(black)}
 \makeatother

\begin{document}

\title{Skyrmion Superfluidity in Two-Dimensional Interacting Fermionic Systems} 

\author{Giandomenico Palumbo${^*}$}
 \affiliation{School of Physics and Astronomy, University of Leeds, Leeds, LS2 9JT, United Kingdom}
 \author{Mauro Cirio${^*}$}
 \affiliation{Interdisciplinary Theoretical Science Research Group (iTHES), RIKEN, Wako-shi, Saitama 351-0198, Japan}

\date{\today}


\begin{abstract}
In this article we describe a multi-layered honeycomb lattice model of interacting fermions which supports a new kind of parity-preserving skyrmion superfluidity. We derive the low-energy field theory  describing a non-BCS fermionic superfluid phase by means of functional fermionization. Such effective theory is a new kind of non-linear sigma model, which we call double skyrmion model. In the  bi-layer case, the quasiparticles of the system (skyrmions) have bosonic statistics and replace the Cooper-pairs role. Moreover, we show that the model is also equivalent to a Maxwell-BF theory, which naturally establishes an effective Meissner effect without requiring a breaking of the gauge symmetry. Finally, we map effective superfluidity effects to identities among fermionic observables for the lattice model. This provides  a signature of our theoretical skyrmion superfluidy that can be detected in a possible implementation of the lattice model in a real quantum system.
\end{abstract}

\maketitle

{\bf Introduction.--} Quantum field theory (QFT) plays a fundamental role in the description of strongly correlated systems and topological phases of matter. For example, free and self-interacting relativistic fermions  emerging in condensed matter systems can be described by  Dirac and Thirring theories respectively \cite{Ando, Qi1,Palumbo1,Palumbo2,Cirio}. At the same time, the ground states of fractional quantum Hall states, topological insulators and superconductors are opportunely described by bosonic topological QFTs like Chern-Simons and BF theories \cite{Fradkin1,Moore1,Hansson}. 
%In the low energy limit, it is possible to show the equivalence between instances of some of these fermionic and bosonic theories by means of functional bosonization techniques \cite{Fradkin1}. 
Another class of bosonic QFT contains the non-linear sigma models (NLSM) which describe the physics of Heisenberg antiferromagnets \cite{Haldane1} and symmetry protected topological phases \cite{Wen1,Xu}. The addiction of a topological term in the theory (Hopf term) \cite{Abanov1} allows for the skyrmions (the quasiparticles present in the model) to acquire fermionic, bosonic or anyonic statistics depending on the value of the coefficient in front of the Hopf term and the value of their topological charge \cite{Wilczek,Polyakov}.
Importantly, bosonic QFTs reveal several features which characterize the physics of superconductivity. In particular, skyrmions have been used to define and describe a parity-breaking two-dimensional non-BCS superconductivity \cite{Abanov2,Senthil1,Baskaran}, while BF theory is a candidate as the effective theory for some strongly correlated fermionic systems \cite{Cirio} and spin Hall states \cite{Senthil1}. BF theory naturally describe the Meissner effect \cite{Hansson,Mavromatos1,Semenoff}, which represents the smoking-gun evidence of superconducting phase. These non-BCS superconducting mechanisms could be used to get insights on the physics of high-temperature superconductors \cite{Mavromatos1, Mavromatos2}.

The goal of this this letter is to provide a new fermionic (multi-layered) honeycomb lattice model that combines characteristics of \emph{both} skyrmions and BF theory in an unified way. This allows us to prove the existence of a parity-preserving non-BCS superfluid phase (analog neutral version of superconducting phase).
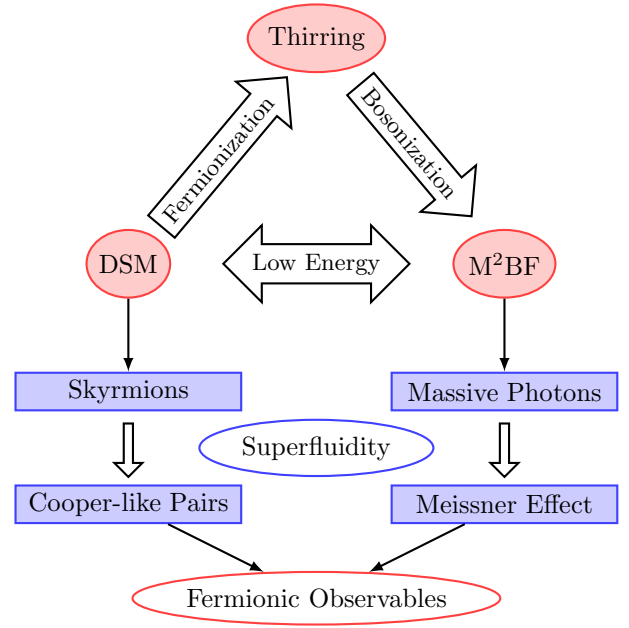
\begin{figure}[h!]
\begin{center}
\begin{tikzpicture}[>=latex,thick,font=\normalsize]

%  \node at (0,0) [draw,rectangle] (a) {Thirring};

  \node at (0,0.2)[ellipse,draw=red!75,fill=red!20,thick,
inner sep=0pt,minimum size=9mm]  (a) {Thirring};

 % \node[inner sep=0,minimum size=0,right of=a] (k) {}; % invisible node
  \node[ellipse,draw=red!75,fill=red!20,thick,
inner sep=0pt,minimum size=9mm]  at (2.5,-2.8) (b) {M${}^2$BF};
  \node at (-2.5,-2.8)[ellipse,draw=red!75,fill=red!20,thick,
inner sep=0pt,minimum size=9mm] (c) {DSM};

  % 1st pass: draw arrows
%\node [fill=blue!50, double arrow, draw=none,rotate =-50] at (1,-1) {arrow 1};
\node [ single arrow, draw,rotate =-50,anchor=center] at  ($(0,0.1)+0.5*($(a)$)+0.5*($(b)$)$) {\small{Bosonization}};
\node [ single arrow, draw,rotate =50] at (-1.30,-1.40) {\small Fermionization};
\node [ double arrow, draw,anchor=center] at  (0,-2.8)  {\small Low Energy}; 
 %\draw[vecArrow] (a) to (b);
%  \draw[vecArrow] (a) to (c);
  %\draw[vecArrow] (b) to (c);
  % 2nd pass: copy all from 1st pass, and replace vecArrow with innerWhite
%  \draw[innerWhite] (a) to (b);
%  \draw[innerWhite] (k) |- (c);

  \node at (-2.5,-4.5)[rectangle,draw=blue!75,fill=blue!20,thick,
inner sep=0pt,minimum width=3cm,minimum height=.5cm] (d) {Skyrmions};
\draw [->] (c) -- ( d);

  \node at (-2.5,-6.0)[rectangle,draw=blue!75,fill=blue!20,thick,
inner sep=0pt,minimum width=3cm,minimum height=.5cm] (dd) {Cooper-like Pairs};
\node [scale=0.5, single arrow, draw,rotate=-90] at (-2.5,-5.25) {~~~~~~~~~~}; 
%\draw [->] (d) -- (dd);

  \node at (2.5,-4.5)[rectangle,draw=blue!75,fill=blue!20,thick,
inner sep=0pt,minimum width=3cm,minimum height=.5cm] (e) {Massive Photons};
\draw [->] (b) -- ( e); 

  \node at (2.5,-6.0)[rectangle,draw=blue!75,fill=blue!20,thick,
inner sep=0pt,minimum width=3cm,minimum height=.5cm] (ee) {Meissner Effect};
\node [scale=0.5, single  arrow, draw,rotate=-90] at (2.5,-5.25) {~~~~~~~~~~}; 

%\node at (0,-5.75) [rectangle,draw,minimum width=8.5cm,minimum height=3cm]{};
\node at (0,-5.25)  [ellipse,draw=blue!75] {Superfluidity};
 % Note: If you have no branches, the 2nd pass is not needed
 
   \node at (0,-7.25)[ellipse,draw=red!75,fill=none,thick,
inner sep=0pt,minimum width=3cm,minimum height=.7cm] (s) {Fermionic Observables};
%\path[line] (dd)--(ss)
\draw [->] (dd) -- (s); 
\draw [->] (ee) -- (s); 

\end{tikzpicture}
\end{center}
\caption{\label{Fig1}Sketch of the logical structure among the effective field theories describing the model and the corresponding physical properties associated to fermionic superfluidity.}
\end{figure}\\
The logical structure of the article is sketched in Fig. \ref{Fig1}. Specifically, the system is described by a fermionic Hubbard-like model which gives rise, in the low-energy limit, to a (2+1)-dimensional chiral-invariant Thirring model \cite{Eguchi} supporting self-interacting Dirac particles. By using functional fermionization techniques \cite{Mavromatos2,Huerta}, we show that this theory is equivalent to a new kind of skyrmion model which is invariant under parity and time-reversal transformations. We call it double skyrmion model (DSM). Interestingly, the statistics of the skyrmions depends on the number of layers. For one layer skyrmions behave like (neutral) fermions and they can pair in order to form Cooper bosons. We focus on bi-layer systems where skyrmions behave as (neutral) bosons and represent the natural Cooper-like pairs in the (fermionic) superfluid phase. In addiction, we show that the system can also be described by a double(Maxwell)-BF (M$^{2}$BF) theory which is a particular instance of a topologically massive gauge theory (TMGT). This equivalence can be shown either by integration of the scalar skyrmionic field  or directly from the fermionic Thirring model by means of functional bosonization \cite{Fradkin2}. In the TMGT theory, effective photons  acquire a mass as a consequence of  topological interactions. This naturally leads to the London equations of superconductivity (fermion superfluidity) \cite{Mavromatos1} which effectively combine Meissner effect and infinite conductivity. We finally show how physical fermionic observables can probe the skyrmion superfluid mechanism described by the model.
%\footnote{* The authors contributed equally to this paper.}

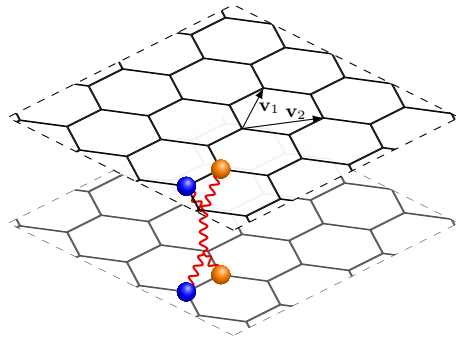
\begin{figure}[t!]
\pgfdeclarelayer{L1}
\pgfdeclarelayer{L2}
\pgfdeclarelayer{L3}
\pgfdeclarelayer{L4}
\pgfdeclarelayer{L5}
\pgfdeclarelayer{L6}
\pgfdeclarelayer{L12}
\pgfsetlayers{L1,L2,L5,L6,L3,L4,L12}
\begin{tikzpicture}[>=latex,scale=.5,every node/.style={draw,shape=regular polygon, regular polygon sides=6, minimum size=.9cm},on grid]
		
    %slanting: production of a set of n 'laminae' to be piled up. N=number of grids.
    \begin{pgfonlayer}{L1}

    \begin{scope}[
            yshift=-83,every node/.append style={
            yslant=0.5,xslant=-1},yslant=0.5,xslant=-1,black!70
            ]
        % opacity to prevent graphical interference
       % \fill[white,fill opacity=0.9] (0,0) rectangle (5,5);
       % \draw[step=4mm, black] (0,0) grid (5,5); %defining grids

 %   \pgfsetfillopacity{0.9};
    \clip (-1,-1) rectangle (5,5);
\node at (0,0) (a1) {};
%\node[anchor=south] at ($0*(a.center)+(a.side 1)$) {D};
%\node[anchor=south] at ($(0,-0.025)+($(a1.side 1)$)$) (b1) {};
\node[anchor=south] at (a1.side 1)(b1) {};
\node[anchor=south] at (b1.side 1)(c1) {};
\node[anchor=south] at (c1.side 1)(d1) {};
\node[anchor=north] at (a1.side 4)(e1) {};

\node[anchor=corner 3] at ($(-0.01,0.000)+($(a1.corner 1)$)$)(a2) {};
\node[anchor=south] at (a2.side 1)(b2) {};
\node[anchor=south] at (b2.side 1)(c2) {};
\node[anchor=south] at (c2.side 1)(d2) {};
%\node[anchor=corner 3] at (e1.corner 1)(e2) {};
\node[anchor=corner 3] at ($(-0.01,0.000)+($(e1.corner 1)$)$)(e2) {};

\node[anchor=corner 3] at ($(-0.01,0.000)+($(a2.corner 5)$)$)(a3) {};
\node[anchor=south] at (a3.side 1)(b3) {};
\node[anchor=south] at (b3.side 1)(c3) {};
\node[anchor=south] at (c3.side 1)(d3) {};
%\node[anchor=corner 3] at (e2.corner 5)(e3) {};
\node[anchor=corner 3] at ($(-0.01,0.000)+($(e2.corner 5)$)$)(e3) {};

%\node[anchor=corner 3] at (a.corner 1)(h) {};

\node[anchor=corner 3] at  ($(-0.01,0.000)+($(a3.corner 1)$)$)(a4) {};
\node[anchor=south] at (a4.side 1)(b4) {};
\node[anchor=south] at (b4.side 1)(c4) {};
\node[anchor=south] at (c4.side 1)(d4) {};
\node[anchor=corner 3] at  ($(-0.01,0.000)+($(e3.corner 1)$)$)(e5) {};

\node[anchor=corner 3] at  ($(-0.01,0.000)+($(a4.corner 5)$)$)(a5) {};
\node[anchor=south] at (a5.side 1)(b5) {};
\node[anchor=south] at (b5.side 1)(c5) {};
\node[anchor=south] at (c5.side 1)(d5) {};
\node[anchor=corner 3] at  ($(-0.01,0.000)+($(e5.corner 5)$)$)(e6) {};

\node[anchor=corner 5] at  ($(0.01,0.000)+($(a1.corner 3)$)$)(a7) {};
\node[anchor=south] at (a7.side 1)(b7) {};
\node[anchor=south] at (b7.side 1)(c7) {};
\node[anchor=south] at (c7.side 1)(d7) {};
\node[anchor=corner 5] at  ($(0.01,0.000)+($(e1.corner 3)$)$)(e7) {};

%\node at (0,1.7) {}; 
%\node at (0,2.55) {}; 
%\node at (.75,0.45){};

%\draw[black,thick,fill=yellow,draw=none] (-1,5) rectangle (5 ,6);
%\draw[black,thick,fill=yellow,draw=none] (-1,-1) rectangle (5 ,-2);
%\draw[black,thick,fill=yellow,draw=none] (5,-2) rectangle (6 ,6);
%\draw[black,thick,fill=yellow,draw=none] (-2,-2) rectangle (-1 ,6);

\draw[black,thick,fill=white,draw=none] (-1,5) rectangle (5 ,6.42);
\draw[black,thick,fill=white,draw=none] (-1,-1) rectangle (5 ,-2.42);
\draw[black,thick,fill=white,draw=none] (5,-2.42) rectangle (6.42 ,6.42);
\draw[black,thick,fill=white,draw=none] (-2.42,-2.42) rectangle (-1 ,6.42);

      % \draw[step=1mm, red!50,thin] (3,1) grid (4,2);  %Nested Grid
        \draw[black!75, dashed] (-1,-1) rectangle (5,5);%marking borders
% \draw[black, thick] (-2.42,-2.42) rectangle (6.42,6.42);%marking borders

        %\fill[red] (0.05,0.05) rectangle (0.35,0.35);
        %Idem as above, for the n-th grid:
    \end{scope}

%\node[anchor=center,draw=none,scale=0.12] at (a1.corner 1){\pgfdeclareradialshading{sphere}{\pgfpoint{0.5cm}{0.5cm}}%
%{rgb(0cm)=(0.9,0,0);
%rgb(0.7cm)=(0.7,0,0);
%rgb(1cm)=(0.5,0,0);
%rgb(1.05cm)=(1,1,1)}
%\pgfuseshading{sphere}
%};

%\node[anchor=center,draw=none,scale=0.12] at (a1.corner 2){\pgfdeclareradialshading{sphere}{\pgfpoint{0.5cm}{0.5cm}}%
%{rgb(0cm)=(0,0,0.9);
%rgb(0.7cm)=(0,0,0.7);
%rgb(1cm)=(0,0,0.5);
%rgb(1.05cm)=(1,1,1)}
%\pgfuseshading{sphere}
%};

     \end{pgfonlayer} ;
             \begin{pgfonlayer}{L12}
% Sphere
\shade [ball color=orange] (a1.corner 1) circle (.25cm);% Sphere
\shade [ball color=blue] (a1.corner 2) circle (.25cm);

\coordinate (A) at  (a1.corner 1);
\coordinate (B) at  (a1.corner 2);

        \end{pgfonlayer}{L12}

    \pgfsetfillopacity{0.93};

    \begin{scope}[
            yshift=-03,every node/.append style={
            yslant=0.5,xslant=-1},yslant=0.5,xslant=-1,black
            ]
        % opacity to prevent graphical interference
       % \fill[white,fill opacity=0.9] (0,0) rectangle (5,5);
       % \draw[step=4mm, black] (0,0) grid (5,5); %defining grids
%\node at (0,0){};
%\node at (0,.85) {};
%\node at (0,1.7) {}; 
%\node at (0,2.55) {}; 
%\node at (.75,0.45){};

   %  \draw[thick,fill=yellow] (-1,-1) rectangle (5,5);

               \begin{pgfonlayer}{L2}
               
    \pgfsetfillopacity{0.93};
       \draw[black, dashed,fill=white] (-1,-1) rectangle (5,5);%marking borders
       \end{pgfonlayer}
\begin{pgfonlayer}{L3}
    \clip (-1,-1) rectangle (5,5);
\node at (0,0) (a1) {};
%\node[anchor=south] at ($0*(a.center)+(a.side 1)$) {D};
%\node[anchor=south] at ($(0,-0.025)+($(a1.side 1)$)$) (b1) {};
\node[anchor=south] at (a1.side 1)(b1) {};
\node[anchor=south] at (b1.side 1)(c1) {};
\node[anchor=south] at (c1.side 1)(d1) {};
\node[anchor=north] at (a1.side 4)(e1) {};

\node[anchor=corner 3] at ($(-0.01,0.000)+($(a1.corner 1)$)$)(a2) {};
\node[anchor=south] at (a2.side 1)(b2) {};
\node[anchor=south] at (b2.side 1)(c2) {};
\node[anchor=south] at (c2.side 1)(d2) {};
%\node[anchor=corner 3] at (e1.corner 1)(e2) {};
\node[anchor=corner 3] at ($(-0.01,0.000)+($(e1.corner 1)$)$)(e2) {};

\node[anchor=corner 3] at ($(-0.01,0.000)+($(a2.corner 5)$)$)(a3) {};
\node[anchor=south] at (a3.side 1)(b3) {};
\node[anchor=south] at (b3.side 1)(c3) {};
\node[anchor=south] at (c3.side 1)(d3) {};
%\node[anchor=corner 3] at (e2.corner 5)(e3) {};
\node[anchor=corner 3] at ($(-0.01,0.000)+($(e2.corner 5)$)$)(e3) {};

%\node[anchor=corner 3] at (a.corner 1)(h) {};

\node[anchor=corner 3] at  ($(-0.01,0.000)+($(a3.corner 1)$)$)(a4) {};
\node[anchor=south] at (a4.side 1)(b4) {};
\node[anchor=south] at (b4.side 1)(c4) {};
\node[anchor=south] at (c4.side 1)(d4) {};
\node[anchor=corner 3] at  ($(-0.01,0.000)+($(e3.corner 1)$)$)(e5) {};

\node[anchor=corner 3] at  ($(-0.01,0.000)+($(a4.corner 5)$)$)(a5) {};
\node[anchor=south] at (a5.side 1)(b5) {};
\node[anchor=south] at (b5.side 1)(c5) {};
\node[anchor=south] at (c5.side 1)(d5) {};
\node[anchor=corner 3] at  ($(-0.01,0.000)+($(e5.corner 5)$)$)(e6) {};

\node[anchor=corner 5] at  ($(0.01,0.000)+($(a1.corner 3)$)$)(a7) {};
\node[anchor=south] at (a7.side 1)(b7) {};
\node[anchor=south] at (b7.side 1)(c7) {};
\node[anchor=south] at (c7.side 1)(d7) {};
\node[anchor=corner 5] at  ($(0.01,0.000)+($(e1.corner 3)$)$)(e7) {};

\draw [->] (a2.corner 1) -- ( c3.corner 5); 
\draw [->] (a2.corner 1) -- ( b3.corner 5); 

 \end{pgfonlayer} ;

%    \shade (a1.corner 1) circle (.3cm);
     %   \draw[black, dashed,fill=white] (-1,-1) rectangle (5,5);%marking borders

      % \draw[step=1mm, red!50,thin] (3,1) grid (4,2);  %Nested Grid
    %    \draw[black, dashed,fill=white] (-1,-1) rectangle (5,5);%marking borders
        %\fill[red] (0.05,0.05) rectangle (0.35,0.35);
        %Idem as above, for the n-th grid:
    \end{scope}
    
        \pgfsetfillopacity{1};
    
    %\pgfsetfillopacity{0.13};
    
     \begin{pgfonlayer}{L4} ;

%    \node[anchor=center,draw=none,scale=0.12] at (a1.corner 1){\pgfdeclareradialshading{sphere}{\pgfpoint{0.5cm}{0.5cm}}%
%{rgb(0cm)=(0.9,0,0);
%rgb(0.7cm)=(0.7,0,0);
%rgb(1cm)=(0.5,0,0);
%rgb(1.05cm)=(1,1,1)}
%\pgfuseshading{sphere}
%};
%
%\node[anchor=center,draw=none,scale=0.12] at (a1.corner 2){\pgfdeclareradialshading{sphere}{\pgfpoint{0.5cm}{0.5cm}}%
%{rgb(0cm)=(0,0,0.9);
%rgb(0.7cm)=(0,0,0.7);
%rgb(1cm)=(0,0,0.5);
%rgb(1.05cm)=(1,1,1)}
%\pgfuseshading{sphere}
%};
    
    % Sphere
\shade [ball color=orange] (a1.corner 1) circle (.25cm);% Sphere
\shade [ball color=blue] (a1.corner 2) circle (.25cm);
    
    \node[anchor=center,draw=none,scale=0.8] at ($(0.43,0.0)+0.5*(c3.corner 5)+0.5*($(a2.corner 1)$)$){{${\bf v}_1$}};
\node[anchor=center,draw=none,scale=0.8] at ($(0.35,0.25)+0.5*(b3.corner 5)+0.5*($(a2.corner 1)$)$){{${\bf v}_2$}};

 \end{pgfonlayer} ;

            \begin{pgfonlayer}{L5}
    
%    \draw [-,decorate,decoration=snake,rotate=90,thick,red!90!black] ($0.5*($(a1.corner 1)$)+0.5*($(a1.corner 2)$)$) -- ($0.5*($(A)$)+0.5*($(B)$)$);
    
        \draw [-,decorate,decoration={snake,amplitude=.5mm,segment length=1.5mm},rotate=90,thick,red!90!black] ($0.5*($(a1.corner 1)$)+(-0.9,0)+0.5*($(a1.corner 2)$)$) -- ($0.5*($(A)$)+(0.8,0)+0.5*($(B)$)$);

                \draw [-,decorate,decoration={snake,amplitude=.5mm,segment length=1.5mm},rotate=90,thick,red!90!black] ($0.5*($(a1.corner 1)$)+(-0.9,0)+0.5*($(a1.corner 2)$)$) -- (a1.corner 1);
                
                 \draw [-,decorate,decoration={snake,amplitude=.5mm,segment length=1.5mm},rotate=90,thick,red!90!black] ($0.5*($(a1.corner 1)$)+(-0.9,0)+0.5*($(a1.corner 2)$)$) -- (a1.corner 2);
                 
             \end{pgfonlayer}
             
                         \begin{pgfonlayer}{L6}
        \draw [-,decorate,decoration={snake,amplitude=.5mm,segment length=1.5mm},rotate=90,thick,red!90!black] (A)-- ($0.5*($(A)$)+(0.8,0)+0.5*($(B)$)$);
        
        \draw [-,decorate,decoration={snake,amplitude=.5mm,segment length=1.5mm},rotate=90,thick,red!90!black] (B)-- ($0.5*($(A)$)+(0.8,0)+0.5*($(B)$)$);

            \end{pgfonlayer}

\end{tikzpicture}
\caption{\label{Fig2}Tight binding for $n=2$. Fermions hop along the edges of two honeycomb lattice layers as described by the Hamiltonian in Eq. (\ref{eq:H0}).
For each spin species and each layer the unit cell contains two fermions: $a$ (orange) and $b$ (blue). Fermions in different layers but in unit cells with same in-layer coordinates  interact through current-current interactions (wavy line) as described by the Hamiltonian in Eq. (\ref{eq:HI}).}
\end{figure}

{\bf Lattice Model.--} 
\noindent We consider $n$ two-dimensional layers of spinful fermions stacked one on the top of each other (Fig. \ref{Fig2}).  Within each layer fermions are localized on a honeycomb lattice. In the case $n=1$ the fermion hopping  is described by the following graphene-like (spin $s=\uparrow, \downarrow$ dependent) Hamiltonian
%\begin{equation}
%\label{eq:H0}
%H_0=\bar{H}_0^\uparrow+\bar{H}_0^\downarrow\;\;,
%\end{equation}
%where
\begin{equation}\label{eq:H0}
\begin{array}{lll}
H_0&=&\pm c \sum_{s, r} \left[(a^{s\dagger}_r b^s_{r+{\bf v}_1}+a^{s\dagger}_r b^s_{r+{\bf v}_2}+a^{s\dagger}_r b^s_{r})\right.\\
&&\left.+mc( a^{s\dagger}_r a^s_{r}- b^{s\dagger}_r b^s_{r})\right]\;\;.
\end{array}
\end{equation}
Here, the overall sign depends on the orientation of the spin  and $a_r$ and $b_r$ are the fermion operators at position ${\bf r}\in \Lambda$ where $\Lambda=n_1 {\bf v}_1+n_2 {\bf v}_2$ is the lattice of unit cells of the model ($n_1,n_2\in\mathbb{N}$ and ${\bf v}_1=(\frac{\sqrt{3}}{2},\frac{3}{2})$ and  ${\bf v}_2=(-\frac{\sqrt{3}}{2},\frac{3}{2})$. 
This Hamiltonian describes hopping terms along the links of the honeycomb lattice with a real tunneling coefficient $c$ and a staggered chemical potential (with energy scale $mc^2$) and it can be exactly solved. The spectrum  becomes gapless at two independent points  $P_\pm$ in momentum space. The low energy physics around these points is effectively described by a standard massive Dirac Hamiltonian
\begin{equation}
H_0=\sum_{\pm}\int d^2 k \left[\Psi_\pm^\dagger(c k_x\alpha^x+c k_y\alpha^y+m c^{2}\beta)\Psi_\pm\right]\;\;,
\end{equation}
where the matrices $\alpha$ and $\beta$ belong to an euclidean Clifford algebra and where, for clarity, the energy scales $c$ and $m c^2$ have been renormalized (for details, see appendix). The spinors $\Psi_\pm$ depend on the momentum space coordinate $k$ as $\Psi_\pm=\left(\begin{array}{lllll}a^\uparrow_\pm(k^\pm) & b^\uparrow_\pm(k^\pm)& a^\downarrow_\pm(k^\pm)& b^\downarrow_\pm(k^\pm)\end{array}\right)^T$ where $a_\pm$ are the Fourier transformed fermion operators evaluated at the Fermi points $P_\pm$ respectively and where $k^+=(k_x,k_y)$ and $k^-=(-k_x,k_y)$. Note that it is possible to induce the same mass term in the above Hamiltonian by replacing the staggered chemical potential in (\ref{eq:H0}) with a standard Haldane term \cite{Haldane2}.
\\
We now consider the general case of $n$ such layers (we will be mainly interested in the case $n=2$) and label their free Hamiltonians by $j=1,\dots,n$ so that $H_0\rightarrow H_0^j$. To connect the layers we add current-current interactions to the free model
\begin{equation}
\label{eq:HI}
\begin{array}{lll}
H=\sum_{j=1}^n H_0^j + H_I\;\;,
\end{array}
\end{equation}
with $H_I=\frac{g^2}{2}[(\sum_j J^j_\mu)^2+(\sum_j J^{j5}_\mu)^2]$, where $\mu=0,1,2$, where the spinor and the currents are, respectively,  $\Psi_j=\left(\begin{array}{lllll}a^{j\uparrow} & b^{j\uparrow}& a^{j\downarrow}& b^{j\downarrow}\end{array}\right)^T$, $J^{j\mu}=\bar{\Psi}_j\gamma^\mu\Psi_j$ and $J^{j5\mu}=\bar{\Psi}_j\gamma^5\gamma^\mu\Psi_j$, and where the $\gamma$s are Dirac gamma matrices.  In the case of a single layer, we have that
\begin{equation}
H_I=3 g^{2}\left[|(a^{\uparrow}b^{\uparrow}+a^{\downarrow}b^{\downarrow})|^{2}+|(a^{\uparrow}b^{\downarrow}-a^{\downarrow}b^{\uparrow})|^{2}\right]\;\;.
\end{equation}
The less compact, but similar, expression for the case $n=2$ can be found in the Supplemental Material \cite{SM}. Around each Fermi point $P_\pm$ the low-energy effective physics is described by the following partition function $Z_F=\int \mathcal{D}\bar{\Psi}_j  \mathcal{D}\Psi_j e^{\frac{i}{\hbar} S_F}$,
%\begin{equation}
%\label{ZF}
%Z_F=\int \mathcal{D}\bar{\Psi}_j  \mathcal{D}\Psi_j e^{\frac{i}{\hbar} S_F}\;\;,
%\end{equation}
where $S_F=S_0+S_I$ with
\begin{equation}
\begin{array}{lll}
S_0&=&\sum_{j=1}^n\int d^3 x\bar{\Psi}_j(c i \slashed{\partial}-mc^2)\Psi_j\\
S_I&=&\int d^3 x\frac{g^2}{2}[(\sum_{j=1}^n J^j_\mu)^2+(\sum_{j=1}^n J^{j5}_\mu)^2]\;\;.
\end{array}
\end{equation}
This model is nothing but a (generalized) chiral-invariant Thirring model \cite{Eguchi}. In the following we will work in units such that $c=\hbar=1$ and without losing of generality we will consider the physics only around one Fermi point.

%%%%%%%%%%%%%%%%%%%%%%%%%%%%%
{\bf Double skyrmion model and functional fermionization.--} We now introduce a double skyrmion model (DSM) which is a double O(3)-Hopf non-linear sigma model 
\begin{equation}
\label{eq:O3Hopff}
S_{(\text{O}(3)-\text{H})^2}=\sum_{t=\pm}\int d^3 x\left[\frac{1}{2g_0^2}(\partial_\mu{\bf m}^t\partial^\mu {\bf m}^t)+t n\pi H^t\right]\;\;,
\end{equation}
where $n$ coincides with the number of layers, the fields ${\bf m}^\pm\in\mathbb{R}^3$ satisfy the non-linear constraint $|{\bf m}^\pm|^{2}=1$ and the two Hopf terms $H^{t}$ are topological invariants \cite{Abanov1}. Due to the different sign in front of the Hopf terms, this theory describes independent skyrmions and anti-skyrmions which have opposite values of the topological charges $Q^+_T=-Q^-_T$ which assume only integer values \cite{SM}.
%and are given by $Q^\pm_T=\int d^2 x J^{\pm0}_S$, where $J^{\pm0}_S$ are the 0th components of the two skyrmion currents.
Each (anti-) skyrmion has a spin $S$ given by \cite{Karabali}
\begin{equation}
\begin{array}{lll}
S=(n /2)Q^{\pm 2}_T\;.
\end{array}
\end{equation}
%where $|Q_T^{\pm}|=1$ represents the simplest {\color{red} I think it is necessary to specify what simplest means here} topological configuration.
This shows that, depending on the number of layers, the statistics of the skyrmions can be either bosonic or fermionic. In particular, for a bi-layer systems (anti-) skyrmions behave like bosons and in our context take the role Cooper-like pairs.\\
Following \cite{Mavromatos2}, we now want to show that the partition function of this bosonic theory is equivalent to that one of the chiral-invariant Thirring theory by employing functional \emph{fermionization}.
 Let us start by defining the equivalent CP form \cite{Fradkin1} of the O(3) NLSM in Eq.  (\ref{eq:O3Hopff}) in which the Hopf terms are recast as Chern-Simons terms. We will refer to it as a double CP-Chern-Simons (CP-CS)${}^2$ model
\begin{equation}
\label{eq:2CPCSS}
Z_{\text{(CP-CS)}^2}=\int \mathcal{D}A^+\mathcal{D}A^-\mathcal{D}z^+ \mathcal{D}z^-  e^{i S_{\text{(CP-CS)}^2}}\;\;,
\end{equation}
where
\begin{equation}
\begin{array}{lll}
S_{\text{(CP-CS)}^2}&=& \sum_{t=\pm}\int d^3 x\left.[ \pm \frac{n}{4\pi}\epsilon^{\lambda\mu\nu}A^t_\lambda\partial_\mu A^t_\nu\right.\\
&&\left.+\frac{1}{{g^2}}|(i\partial_\mu-A^t_\mu)z^t|^2\right.]\;.
\end{array}
\end{equation}
Here, $z^\pm=\left(\begin{array}{l}z^\pm_1~~z^\pm_2\end{array}\right)^T$ with the fields $z^\pm_1,z^\pm_2\in\mathbb{C}$ such that $|z^\pm|^2=z^\pm_1z^{\pm*}_1+z^\pm_2z^{\pm*}_2=1$ and $A_{\mu}^{\pm}=-(i/2)z^{\pm*}\partial_{\mu}z^\pm$. 
We now proceed with the fermionization \cite{SM}. The fermions appear quite naturally. In fact, we begin by noticing that, by changing variables to $A_{\mu}=\frac{1}{2}(A^+_{\mu}+A^-_{\mu})$, $B_{\mu}=\frac{1}{2}(A^+_{\mu}-A^-_{\mu})$, the difference of the  two Chern-Simons terms transform into the BF term $\frac{n}{\pi}\int d^3 x \epsilon^{\mu\nu\lambda} B_{\mu}\partial_{\nu}A_{\lambda}$ leading to a (double CP)-BF theory so that $Z_{\text{CP${^2}$-BF}}=Z_{\text{(CP-CS)}^2}$. The BF term can now be ``linearized'' by introducing $n$ fermion species $\chi_{j}$ \cite{Mavromatos1}
%we in fact get a double CP-BF theory described by
%\begin{equation}
%\begin{array}{lll}
%Z_{\text{(CP)}^2- BF}&=&\int \mathcal{D}A\mathcal{D}B\mathcal{D}z^+ \mathcal{D}z^-\\
%&&\exp\{i \int d^3 x\left[  \frac{n}{\pi}BdA\right.\\
%&&\left.+\sum_{t=\pm}\frac{1}{{g^2}} |(i\partial_\mu-(A_\mu+t B_\mu))z^t|^2\right]\}\;\;,
%\end{array}
%\end{equation}
%%%%%%%%%%%%%%%%%%%%%%%%%%%%%%%%%%%%%%%
%Now, we can introduce $n$ species of fermions in order to ``linearize'' the BF term:
%\begin{equation}
%\label{eq:linearization}
%\begin{array}{lll}
%e^{i\frac{n}{\pi}\int d^3 x BdA}&=&\int \mathcal{D}\Psi\mathcal{D}\bar{\Psi}\text{exp}\left[i\sum_j^n(\bar{\Psi}_j (i\slashed{\partial}-m)\Psi_j\right.\\
%&&\left.-\sqrt{2}A_\mu J_\Psi^{j\mu}-\sqrt{2}B_\mu J_\Psi^{5j\mu})\right]
%\end{array}
%\end{equation}
leading to the following intermediate partition function
\begin{equation}
\label{eq:middlee}
\begin{array}{lll}
Z=\int \mathcal{D}A\mathcal{D}B\mathcal{D}z^+ \mathcal{D}z^-\mathcal{D}\chi\mathcal{D}\bar{\chi}\\
\exp\{i  \sum_j\int d^3 x\left.[\bar{\chi}_j (i\slashed{\partial}-m)\chi_j\right.
\left.-\sqrt{2}A_\mu J_\chi^{j\mu}-\sqrt{2}B_\mu J_\chi^{5j\mu})\right.\\
\left.+\frac{1}{{g^2}}\sum_{t=\pm}|(i\partial_\mu-(A_\mu+t B_\mu))z^t|^2\right.]\}\;.
\end{array}
\end{equation}
For each value of the sign, the variable $z^\pm$ can be thought as specifying a coordinate system in a SU(2) algebra via the identification $z^\pm\rightarrow e^{i\xi^\pm_j\sigma^j}$, where $\sigma^j$  are the Pauli matrices and $\xi^\pm_j$ are some fields \cite{SM}.
%\begin{equation}
%\begin{array}{lll}
%z^\pm&=&e^{i\xi^\pm_j\sigma^j}\\
%A_\mu&\rightarrow&A_\mu\sigma^3\\
%B_\mu&\rightarrow&B_\mu\sigma^3\\
%\end{array}
%\end{equation}
Moreover, the gaussian integral over the fields $A$ and $B$ can be easily computed. This cause the fields $\xi_j^\pm$ to effectively decouple and changing the fermionic variables $\chi_{j}\rightarrow \Theta\Psi_{j}$ through a suitable phase $\Theta$ \cite{SM}, we are left with
% As a final result we get
%\begin{equation}
%\begin{array}{lll}
%Z
%&=&\int \mathcal{D}\xi^+ \mathcal{D}\xi^-\mathcal{D}\chi\mathcal{D}\bar{\chi}\\
%&&\exp\{i \int d^3 x\left[ \sum_j(\bar{\chi}_j (i\slashed{\partial}-m)\chi_j\right.\\
%&&+\frac{1}{{g^2}}\int d^3 x \frac{1}{2}(\partial_\mu \xi^A_j)^2+\frac{1}{2}(\partial_\mu \xi^B_j)^2\\
%&&\frac{{g^2}}{2}(\sum_j J^{\chi j}_\mu)^2+\frac{{g^2}}{2}(\sum_j J^{5\chi j}_\mu)^2\\
%&&-(\frac{1}{4{g^2}})^2J^{3A}_\mu-(\frac{1}{4{g^2}})^2J^{3B}_\mu
%\end{array}
%\end{equation}
%where $\xi^A_j=\xi^+_j+\xi^-_j$,  $\xi^B_j=\xi^+_j-\xi^-_j$, $J^{\xi^A 3}_\mu=2\partial_\mu\xi_j^A$, $J^{\xi^B 3}_\mu=2\partial_\mu\xi_j^B$ and $\Psi=e^{i \sqrt{2} (\xi^{3A}+ \gamma^5 \xi^{3B})}\chi$.\\
%We finally see that the fields $\xi$ do  not interact with the fermion so that we can integrate them out to get:
\begin{equation}
\label{eq:ThirringAfterFermionizationn}
\begin{array}{lll}
Z_F=\int \mathcal{D}\Psi\mathcal{D}\bar{\Psi}
\exp\{i \sum_j\int d^3 x\left[\bar{\Psi}_j (i\slashed{\partial}-m)\Psi_j\right.\\
\hspace{2.0cm}\left.+\frac{{g^2}}{2}(J^{ j}_\mu)^2+\frac{{g^2}}{2}(J^{5 j}_\mu)^2\right.]\}\;,
\end{array}
\end{equation}
which is in fact the original chiral-invariant Thirring model introduced in the previous section. As a final comment, we note that, alternatively, it is possible to show that the this fermionic model can be \emph{bosonized} to get the  (CP-CS)${^2}$ model \cite{SM}. 
%%%%%%%%%%%%%%%%%%%%%%%%%%%%%%

{\bf \bf London action.--} 
In this section we show that the effective theory described in Eq. (\ref{eq:2CPCSS}) is equivalent to the London action which effectively describes the physics of superconductivity.\\
In \cite{Mavromatos1,Schakel} it is proven that (at low energy) a Maxwell theory is equivalent to a CP model. We can use this to map the effective theory in Eq. (\ref{eq:2CPCSS}) to a (double Maxwell)-BF theory (M${^2}$BF) \cite{Mavromatos1,Hansson} with action
\begin{equation}
\label{eq:MMBF}
\begin{array}{lll}
S_{\text{M}^2-BF}&=&\int d^3 x\left[  \frac{n}{\pi}\epsilon^{\lambda\mu\nu}B_\lambda\partial_\mu A_\nu-\frac{1}{4 e^2}F_{\mu\nu}(A)F^{\mu\nu}(A)\right.\\
&&\left.-\frac{1}{4 e^2}F_{\mu\nu}(B)F^{\mu\nu}(B)\right]\;\;,
\end{array}
\end{equation}
%
%
%
%\begin{equation}
%Z_{CP}=Z_M
%\end{equation}
%where
%\begin{equation}
%Z_{CP}=\int\mathcal{D}A\mathcal{D}z\mathcal{D}z^\dagger\delta(z^\dagger z-1)e^{\frac{i}{g_0^2}\int d^3 x |(i\partial_\mu-A_\mu)z|^2}
%\end{equation}
%and:
%\begin{equation}
%Z_M=\int \mathcal{D}A e^{-\frac{i}{4e^2}\int d^3 x F(A)_{\mu\nu}F(A)^{\mu\nu}}\;\;,
%\end{equation}
where $F_{\mu\nu}$ is the field strength tensor and the two scales $g_0$ and $e$ are related as showed in \cite{SM}.
%by $i g_0^2\int \frac{d^3 k}{(2\pi)^3)}\frac{1}{k^2-e^2}=1$. The charge $e$ is renormalized in relation to a momentum cutoff $\Lambda$ as $|e|=\frac{(2\pi)^2}{s g_0^2}$ where $\Lambda=\sqrt[3]{\frac{3}{2}}~s ~M$ with $s\ll 1$(low kinetic energy limit).\\
% By using this mapping we can map our double CP-CS to a BF-double-Maxwell:
%\begin{equation}
%\begin{array}{lll}
%S_{(\text{M-BF})^2}&=&\int d^3 x\left[  \frac{n}{4\pi}\epsilon^{\lambda\mu\nu}A^+_\lambda\partial_\mu A^+_\nu- \frac{n }{4\pi}\epsilon^{\lambda\mu\nu}A^-_\lambda\partial_\mu A^-_\nu\right.\\
%&&\left.-\frac{1}{4 e^2}\int d^3 x F_{\mu\nu}F^{\mu\nu}\right]\;\;,
%\end{array}
%\end{equation}
%By defining new fields $A$ and $B$ as $A^+_\mu=A^\mu+B^\mu$ and $A^-_\mu=A^\mu-B^\mu$, we get the dual theory with partition function:
%\begin{equation}
%\label{eq:bosonicTheory}
%\begin{array}{lll}
%Z_{\text{M}^2-BF}&=&\int \mathcal{D}A\mathcal{D}B \\
%&&\text{exp}i\int d^3 x\left[  \frac{n}{\pi}\epsilon^{\lambda\mu\nu}B_\lambda\partial_\mu A_\nu\right.\\
%&&\left.-\frac{1}{4 e^2}F_{\mu\nu}(A)F^{\mu\nu}(A)-\frac{1}{4 e^2}F_{\mu\nu}(B)F^{\mu\nu}(B)\right]\;\;,
%\end{array}
%\end{equation}
We now follow \cite{Mavromatos1} which shows that this theory is equivalent to the London partition function:
\begin{equation}
\label{eq:phiTheory}
Z_\phi=\int \mathcal{D}A\mathcal{D}\phi ~e^{i\int d^3 x\left[ -\frac{1}{4 e^2}F_{\mu\nu}(A)F^{\mu\nu}(A)+2 e^2 (\partial_\mu\phi-\frac{n}{2\pi}A_\mu)^2 \right]}.
\end{equation}
We can see that the (2+2) degrees of freedom of the massless fields $A$ and $B$ are mapped to the (3+1) degrees of freedom of a massive bosonic field $A$ and a massless scalar field $\phi$ which, in this sense, represents a kind of Goldstone boson. The present mechanism, however, does not have any local order parameter like in ordinary BCS theory.
The charge and currents associated with the field $A$ are
\begin{equation}
\label{eq:rhoJJ}
\begin{array}{lll}
\rho=\frac{\delta \mathcal{L}}{\delta A_0}\hspace{1.0cm}
{\bf{J}}_\text{em}=\frac{\delta \mathcal{L}}{\delta \bf{A}}\;\;,
\end{array}
\end{equation}
%Now we have that:
%\begin{equation}
%\rho=\frac{\delta \mathcal{L}}{\delta (\partial_0 \tilde{\phi})}\\
%\end{equation}
%where $\tilde{\phi}$ is such that $\partial_0\tilde{\phi}=\partial_0-\frac{n}{2\pi}A_0$ so that $\tilde{\phi}$ and $A_0$ are conjugate variables. The Hamilton equation of motion are:
%\begin{equation}
%\partial_0\tilde{\phi}=\frac{\delta\mathcal{H}}{\delta \rho}
%\end{equation}
%and this gives:
%\begin{equation}
%\label{eq:Vphi}
%V=\frac{\partial\tilde{\phi}}{\partial t}
%\end{equation}
%where $V$ is the voltage. So, the presence of steady state currents implies 
%\begin{equation}
%\label{eq:V0}
%V=0
%\end{equation}
%Another, maybe more precise, way to put it is to first 
while the effective magnetic and electric fields inside the material are simply given by
\begin{equation}
\label{eq:magneticAndElectricFields2}
\begin{array}{lll}
E^{i}&=&\frac{1}{2}\epsilon^{i\mu\nu}F(A)_{\mu\nu} \hspace{0.9cm}
B_\text{mag}=\frac{1}{2}\epsilon^{0\mu\nu}F(A)_{\mu\nu}\; ,
\end{array}
\end{equation}
where $i=x,y$. The effective physics described by the massive field $A$, implies both  a Meissner  and infinite conductivity effects. In fact, the (effective) magnetic field intensity decays exponentially inside the material (Meissner effect) due to the presence of superficial dissipationless screening currents. In particular we have that
\begin{equation}
\label{eq:Meiss}
B_\text{mag}=0
\end{equation}
in the bulk of the material. As shown in \cite{Mavromatos1} a zero voltage can be defined in the presence of steady currents. These screening currents flow within a penetration depth $\lambda\propto s g^2$ \cite{SM} from the boundary of the material.  In this sense, the system has infinite conductivity $\sigma$ and follows the perfect conductivity relation ${\bf{E}}=\sigma{\bf{J}}_\text{em}$. \\
%while the currents and electric fields are related by Newton`s law as ${\bf{E}}=\Lambda\frac{d}{dt}\bf{J}$ within a penetration depth from the boundary given by $\lambda\propto s g^2$. Alternatively, we can say that the system is dissipationless that is ${\bf{E}}=\sigma\bf{J}$ with $\sigma=\infty$.
%
%The time dependence of the current $\bf{J}=\frac{\delta \mathcal{L}}{\delta \bf{A}}$ is constrained by Newton`s law:
%\begin{equation}
%{\bf{E}}=\Lambda\frac{d}{dt}\bf{J}
%\end{equation}
%We can compare this behaviour to the one the introduction of dissipation which is described by the Drude model as ${\bf{E}}=\sigma\bf{J}$.
%On the other hand he magnetic field is expelled inside the material:
%\begin{equation}
%B=0
%\end{equation}
%{\color{red}{maybe we need to cite a source where these two equations are derived from the London action or explain things more.}}\\
%We now observe that $\pi_\phi$, the momentum conjugate to the variable $\phi$ since $\pi_\phi=\frac{\delta\mathcal{L}}{\delta\partial_0\phi}
%=\frac{2\pi}{n}\frac{\delta\mathcal{L}}{\delta(A_0)}=\frac{2\pi}{n}\rho$. The Hamilton equation of motion are $\partial_0\phi=\frac{\delta\mathcal{H}}{\delta \rho}$, and this, following Weinberg,  gives $V=\frac{\partial\phi}{\partial t}$ where $V$ is the voltage. So, the presence of steady state currents implies 
%\begin{equation}
%\label{eq:V0}
%V=0
%\end{equation}\\
%%%%%%%%%%%%%%%%%%%%%%%%%%%%%%%

{\bf Fermionization rules and physical observables.--} 
The aim of this section is to map the effective superfluidity physics which describes the model to fermionic observables.
To this end,  we  introduce a minimal coupling interaction with two external fields $A_\text{ext}$ and $B_\text{ext}$ to the fermionic Lagrangian density in Eq. (\ref{eq:ThirringAfterFermionizationn}) via a minimal coupling $\sum_j J^{j}_\mu A^{\mu}_\text{ext}+\sum_j J^{5 j}_\mu B^{\mu}_\text{ext}$ so that $Z_F\rightarrow Z_F(A_\text{ext}, B_\text{ext})$.
%\begin{equation}
%\label{eq:ThirringAfterFermionization2}
%Z_F(A_\text{ext}, B_\text{ext})=e^{i S_F(J^\chi,J^{5\chi})}
%\end{equation}
%where:
%\begin{equation}
%S_F(A_\text{ext}, B_\text{ext})=S_F+\int d^3 x\left[\sum_j J^{\chi j}_\mu A^{\mu}_\text{ext}+\sum_j J^{5\chi j}_\mu B^{\mu}_\text{ext}\right]
%\end{equation}
We can then track these new terms as we follow back all the steps that lead us from Eq. (\ref{eq:middlee}) to Eq. (\ref{eq:ThirringAfterFermionizationn}). The additional terms only cause a shift in the Dirac operator $i\slashed{\partial}-m\rightarrow i\slashed{\partial}-m+\slashed{A}_{\text{ext}}+\gamma^5 \slashed{B}_{\text{ext}}$ which leads to the equivalence $Z_F(A_\text{ext}, B_\text{ext})=Z_{\text{CP}^2BF}(A_\text{ext}, B_\text{ext})$, where  this last partition function 
%$Z_{\text{CP}^2BF}(A_\text{ext}, B_\text{ext})$  
has an additional term $\frac{n}{\sqrt{2}\pi}\int \epsilon^{\mu\nu\lambda}(-B_{\mu}^\text{ext}\partial_{\nu}A_{\lambda}+A_{\mu}^\text{ext}\partial_{\nu}B_{\lambda}+\frac{1}{\sqrt{2}} B_{\mu}^\text{ext}\partial_{\nu}A_{\lambda}^\text{ext})$ in the action \cite{SM}. %\footnote{For simplicity, we use the differential geometry notation  $d \mathcal{A}\equiv\epsilon^{\mu\nu\lambda}\partial_{\nu}\mathcal{A}_\lambda$ where $\mathcal{A}\equiv A_\mu$ is a vector field. Moreover we define $\int\mathcal{B}d\mathcal{A}\equiv\epsilon^{\mu\nu\lambda} \int d^3 xB_\mu\partial_{\nu}\mathcal{A}_\lambda$.}.
%\begin{equation}
%\begin{array}{lll}
%Z_{\text{CP}^2BF}(A_\text{ext}, B_\text{ext})&=&\int \mathcal{D}A\mathcal{D}B\mathcal{D}z^+ \mathcal{D}z^-\\
%&&\exp\{i \int d^3 x\left[  \frac{n}{\pi}BdA\right.\\
%&&+\sum_{t=\pm}\frac{1}{{g^2}} |(i\partial_\mu-(A_\mu+tB_\mu))z^t|^2\\
%&&\left.+\frac{n}{\sqrt{2}\pi}(-B^\text{ext}dA+A^\text{ext}dB)\right.\\
%&&\left.+\frac{n}{2\pi}B^\text{ext}dA^\text{ext}\right]\;\;.
%\end{array}
%\end{equation}
By taking derivatives of the partition functions with respect to the external fields \cite{SM}, this allows us to prove the following ``fermionization'' rules 
\begin{equation}
\begin{array}{lll}
\langle\sum_{j}J_{\mu}^{j}\rangle_F&=&\frac{n}{\sqrt{2}\pi}\langle\epsilon_{\mu\nu\lambda}\partial_{\nu}B_{\lambda}\rangle_{\text{CP}^2BF} \\
\langle\sum_{j}J_{\mu}^{5 j}\rangle_F&=&-\frac{n}{\sqrt{2}\pi}\langle\epsilon_{\mu\nu\lambda}\partial_{\nu}A_{\lambda}\rangle_{\text{CP}^2BF}\; \;,
\end{array}
\end{equation}
%\begin{equation}
%\begin{array}{lll}
%J^\chi&\leftrightarrow&\frac{n}{\sqrt{2}\pi}d B\\
%J^{5\chi}&\leftrightarrow& - \frac{n}{\sqrt{2}\pi}d A
%\end{array}
%\end{equation}
%which hold in the following sense:
%\begin{equation}
%\label{eq:fermionizationRules1}
%\begin{array}{lll}
%< J>_F&=&\frac{n}{\sqrt{2}\pi}<dB>_{\text{CP}^2BF}\\
%<J^5>_F&=&-\frac{n}{\sqrt{2}\pi}<dA>_{\text{CP}^2BF}
%\end{array}
%\end{equation}
where the expectation values $\langle\rangle_F$ and $\langle\rangle_{\text{CP}^2BF}$ are calculated with respect to the ground state of the  fermionic and bosonic theory respectively. \\
Similarly, we can we can map observables of the London theory to those of a double(Maxwell)-BF
by adding and tracking source terms $F_{\mu\nu}(A)\epsilon^{\mu\nu\lambda}J^A_\lambda$ and $F_{\mu\nu}(B)\epsilon^{\mu\nu\lambda}J^B_\lambda$ to the latter theory \cite{SM} so that $Z_{\text{M}^2\text{BF}}\rightarrow Z_{\text{M}^2\text{BF}}(J^A,J^B)$.
%\begin{equation}
%\begin{array}{lll}
%Z_{\text{M}^2\text{BF}}(J)&=&\int \mathcal{D}A\mathcal{D}B \\
%&&\text{e}\left\{i\int d^3 x\left[  \frac{n}{\pi}\epsilon^{\lambda\mu\nu}B_\lambda\partial_\mu A_\nu-\frac{1}{4 e^2}F_{\mu\nu}(A)F^{\mu\nu}(A)\right.\right.\\
%&&\left.\left.-\frac{1}{4 e^2}F_{\mu\nu}(B)F^{\mu\nu}(B)+F_{\mu\nu}(A)\epsilon^{\mu\nu\lambda}J^A_\lambda+F_{\mu\nu}(B)\epsilon^{\mu\nu\lambda}J^B_\lambda\right]\right\}\;\;,
%\end{array}
%\end{equation}
%These additional terms can be thought as a non-minimal coupling of a current $J$ to the field $A,B$. Alternatevely, we can think of it as a minimal coupling $-2 A_\mu \tilde{J}^\mu$ between the field $A$ and a current $\tilde{J}=\epsilon_{\mu\nu\lambda}\partial^\nu J^\lambda$. \\
%It is now possible to  track the source terms through  the steps that show the equivalence between the partition function with action given in Eq. \ref{eq:MMBF} and the one with action given in Eq. \ref{eq:phiTheory} (see appendix). 
This leads to the equivalence $Z_{\text{M}^2\text{BF}}(J^A,J^B)=Z_\phi(J^A,J^B)$ (see \cite{SM} for its explicit expression).
 %where the action associated with $Z_\phi(J^A,J^B)$ has an additional term $F_{\mu\nu}(A)\epsilon^{\mu\nu\lambda}J_\lambda^A$ and $\partial_\mu\phi\rightarrow\partial_\mu\phi +J_\mu^B$ \cite{SM}. 
 %
 %
%\begin{equation}
%\label{eq:equivalenceMBFphi}
%Z_{\text{M}^2\text{BF}}(J)=Z_\phi(J)
%\end{equation}
%with:
%\begin{equation}
%\begin{array}{lll}
%Z_\phi(J^A,J^B)&=&\int \mathcal{D}A\mathcal{D}\phi \\
%&&e^{ i\int d^3 x\left[ -\frac{1}{4 e^2}F_{\mu\nu}(A)F^{\mu\nu}(A)\right]}\\
%&&e^{\left[ 2 e^2 (\partial_\mu\phi-\frac{n}{2\pi}A_\mu+J^B_\mu)^2+F_{\mu\nu}(A)\epsilon^{\mu\nu\lambda}J^A_\lambda \right]}\;\;.
%\end{array}
%\end{equation}
%The aim of this section is to find physical fermionic observables that we can use to observe these superfluidity effects. In particular, we want to relate physical observables to the effective magnetic and electric fields inside the material
%\begin{equation}
%\label{eq:magneticAndElectricFields2}
%\begin{array}{lll}
%E^{j}&=&\frac{1}{2}\epsilon^{j\mu\nu}F(A)_{\mu\nu}\\
%B&=&\frac{1}{2}\epsilon^{0\mu\nu}F(A)_{\mu\nu}\\
%\end{array}
%\end{equation}
%The time dependence of the current $\bf{J}=\frac{\delta \mathcal{L}}{\delta \bf{A}}$ is constrained by Newton`s law:
%\begin{equation}
%{\bf{E}}=\Lambda\frac{d}{dt}\bf{J}
%\end{equation}
%We can compare this behaviour to the one the introduction of dissipation which is described by the Drude model as ${\bf{E}}=\sigma\bf{J}$.
%On the other hand he magnetic field is expelled inside the material:
%\begin{equation}
%B=0
%\end{equation}
We can now use these correspondences to relate the current $(\rho,{\bf J}_\text{em})$ and the fields $(B_\text{mag},{\bf E})$ defined in Eqs. (\ref{eq:rhoJJ}) and (\ref{eq:magneticAndElectricFields2}) to fermionic observables. 
By inspecting Eq. (\ref{eq:rhoJJ}) and the expression of $Z_\phi(J^A,J^B)$, we first  notice that $\rho=\frac{2\pi}{n}\frac{\delta \mathcal{L}}{J_0^B}$ and $J^i_\text{em}=\frac{2\pi}{n}\frac{\delta \mathcal{L}}{\delta J_i^B}$
%\begin{equation}
%\begin{array}{lll}
%\rho&=&\frac{2\pi}{n}\frac{\delta \mathcal{L}}{J_0^B}\\
%J^j&=&\frac{2\pi}{n}\frac{\delta \mathcal{L}}{\delta J_j^B}
%\end{array}
%\end{equation}
which leads to $\langle\rho\rangle_\phi=\frac{2\pi}{n}\frac{\delta Z_{\text{M}^2\text{BF}}}{\delta J_0^B}$ and $\langle J^i_\text{em}\rangle_\phi=\frac{2\pi}{n}\frac{\delta Z_{\text{M}^2\text{BF}}}{\delta J_i^B}$ (where we implicitly impose $J^B=0$, after the derivative has been taken).
%\begin{equation}
%\begin{array}{lllll}
%<\rho>_\phi&=&\frac{2\pi}{n}\frac{\delta Z_\phi}{\delta J_0^B}_{{\Big|}_{J^B=0}}&=&\frac{2\pi}{n}\frac{Z_{\text{M}^2\text{BF}}}{\delta J_0^B}_{{\Big|}_{J^B=0}}\\
%<J^j>_\phi&=&\frac{2\pi}{n}\frac{\delta Z_\phi}{\delta J_j^B}_{{\Big|}_{J^B=0}}&=&\frac{2\pi}{n}\frac{Z_{\text{M}^2\text{BF}}}{\delta J_j^B}_{{\Big|}_{J^B=0}}\\
%\end{array}
%\end{equation}
%where we used Eq. \ref{eq:equivalenceMBFphi}. 
Finally, from the equivalences between London, (double Maxwell)-BF and (double CP)-BF theories we  find the following correspondence
\begin{equation}
\begin{array}{lllll}
\langle\rho\rangle_\phi&=&\langle\epsilon^{0\mu\nu}F_{\mu\nu}(B)\rangle_{\text{M}^2BF}&=&\frac{2\sqrt{2}\pi}{n}\langle\sum_{j}J_0^{j}\rangle_{\text{F}}\\
\langle J^i_\text{em}\rangle_\phi&=&\langle\epsilon^{i\mu\nu}F_{\mu\nu}(B)\rangle_{\text{M}^2BF}&=&\frac{2\sqrt{2}\pi}{n}\langle\sum_{j}J_i^{j}\rangle_{\text{F}}\;\;.
\end{array}
\end{equation}
%This can be rephrased as $<\rho>_\phi=<\epsilon^{\mu\nu 0}F_{\mu\nu}(B)>_{\text{M}^2BF}$ and $<J^j>_\phi=<\epsilon^{\mu\nu j}F_{\mu\nu}(B)>_{\text{M}^2BF}$.
%\begin{equation}
%\begin{array}{lll}
%<\rho>_\phi&=&<\epsilon^{\mu\nu 0}F_{\mu\nu}(B)>_{\text{M}^2BF}\\
%<J^j>_\phi&=&<\epsilon^{\mu\nu j}F_{\mu\nu}(B)>_{\text{M}^2BF}\\\\
%\end{array}
%\end{equation}
%From the equivalence between the double Maxwell BF and double CP BF, the fermionization rules imply the following equivalence between expectation values $<\rho>_\phi=\frac{2\sqrt{2}\pi}{n}<J_0>_{\text{F}}$ and $<J^j>_\phi=\frac{2\sqrt{2}\pi}{n}<J_j>_{\text{F}}$.\\
%\begin{equation}
%\begin{array}{lll}
%<\rho>_\phi&=&\frac{2\sqrt{2}\pi}{n}<J_0>_{\text{F}}\\
%<J^j>_\phi&=&\frac{2\sqrt{2}\pi}{n}<J_j>_{\text{F}}\\\\
%\end{array}
%\end{equation}
Similarly,  the electric and magnetic fields are simply given by $\langle E^i\rangle=\frac{1}{2}\frac{\delta Z_\phi}{\delta J^A_i}$ and $\langle B_\text{mag}\rangle=\frac{1}{2}\frac{\delta Z_\phi}{\delta J^A_0}$, which implies $\langle E^i\rangle_\phi=\frac{\sqrt{2}\pi}{n}\langle\sum_{j}J_i^{5 j}\rangle_{\text{F}}$ and  $\langle B_\text{mag}\rangle_\phi=\frac{\sqrt{2}\pi}{n}\langle\sum_{j}J_0^{5 j}\rangle_{\text{F}}$ after using the fermionization rules. 
%\begin{equation}
%\begin{array}{lll}
%<E^j>&=&\frac{1}{2}\frac{\delta Z_\phi}{\delta J^A_j}\\
%<B>&=&\frac{1}{2}\frac{\delta Z_\phi}{\delta J^A_0}
%\end{array}
%\end{equation}
%which leads to:
%\begin{equation}
%\begin{array}{lllll}
%<E^j>&=&\frac{1}{2}\frac{\delta Z_{\text{M}^2BF}}{\delta J^A_j}&=&\frac{1}{2}<\epsilon^{\mu\nu j}F_{\mu\nu}(A)>_{\text{M}^2BF}\\
%<B>&=&\frac{1}{2}\frac{\delta Z_{\text{M}^2BF}}{\delta J^A_0}&=&\frac{1}{2}<\epsilon^{\mu\nu 0}F_{\mu\nu}(A)>_{\text{M}^2BF}
%\end{array}
%\end{equation}
%Now, thanks to the ferminization rules, we have:
%\begin{equation}
%\begin{array}{lll}
%<B>_\phi&=&\frac{\sqrt{2}\pi}{n}<J^5_0>_{\text{F}}\\
%<E^j>_\phi&=&\frac{\sqrt{2}\pi}{n}<J^5_j>_{\text{F}}\\\\
%\end{array}
%\end{equation}
These finding are summarized in the following table
\begin{center}
\begin{tabular}{|cc|}
\hline
~~Electromagnetic Quantities~~&~~Fermionic Observables~~\\
\hline
$\langle\rho\rangle_\phi$&$\langle\sum_{j}J_0^{j}\rangle_F$\\
$\langle{\bf{J}}_\text{em}\rangle_\phi$&$\langle{\sum_{j}{\bf{J}}^{j}}\rangle_F$\\
$\langle{\bf E}\rangle_\phi$&$\langle{\sum_{j}{\bf {J}}^{5 j}}\rangle_F$\\
$\langle B_\text{mag}\rangle_\phi$&$\langle\sum_{j}J^{5 j}_0\rangle_F$\\
\hline
\end{tabular}
\end{center}
In this way, the effective Meissner effect in Eq. (\ref{eq:Meiss}) can now be written as follows
\begin{equation}
\label{eq:Messnier}
\begin{array}{lll}
\langle\sum_{j}J^{5 j}_0\rangle_F=0\;\;.
\end{array}
\end{equation}
The validity of such a prediction is confirmed by the skyrmionic interpretation of the model.
In fact, Eq. (\ref{eq:Messnier})  can be derived from an alternative dual point of view. We first notice that the expectation value of the skyrmion currents \cite{SM} $J^{\mu\pm}_S=\frac{1}{2\pi}\epsilon^{\mu\nu\lambda}\partial_{\nu}A_{\lambda}^{\pm}$ can be written as $\langle J^{\pm}_S\rangle_{(\text{CP-CS})^2}=\pm\frac{1}{n\sqrt{2}}(\langle\sum_{j}J^{j}\rangle_F\mp\langle\sum_{j}J^{5 j}\rangle_F)$ by using the fermionization rules.
%which connects the skyrmionic currents (which invovles bosons for $n$ even) to the fermionic observables.\\
%
%
%\begin{equation}
%\label{eq:skyrmionsFermions}
%\begin{array}{lll}
%<J^{+\mu}_S>_{(\text{CP-CS})^2}&=&\frac{1}{n\sqrt{2}}(<J^\mu>_F-<J^{\mu 5}>_F)\\
%<J^{-\mu}_S>_{(\text{CP-CS})^2}&=&-\frac{1}{n\sqrt{2}}(<J^\mu>_F+<J^{\mu 5}>_F)
%\end{array}
%\end{equation}
%Now, from Eq. \ref{eq:O3Hopf} we can see that the two skyrmions have opposite topological charge $Q_T$:
%\begin{equation}
%\label{eq:QTplusQTminus}
%Q_T^+=-Q_T^-
%\end{equation}
%Now:
Now, by definition, the topological charges are the spatial integral of the $0$th component of the current $Q^\pm_T=\int d^2 x J^{0\pm}_S$ so that
\begin{equation}
\begin{array}{lll}
0&=&-\frac{n}{\sqrt{2}}(Q_T^++Q_T^-)=-\frac{n}{\sqrt{2}}\int d^2 x\sum_{t=\pm}\langle J^{0 t}_S\rangle_{(\text{CP-CS})^2}\\
&=&\int d^2 x \langle\sum_{j}J^{5 j}_0\rangle_F\;\;,
\end{array}
\end{equation}
consistently with Eq. (\ref{eq:Messnier}). \\
At the same time, as mentioned above, the system supports steady state currents within a penetration depth $\lambda\propto g^2$ distance from the boundary. By tuning the parameter $g$ to allow the fermionization rules to hold, the Drude relation ${\bf{E}}=\sigma\bf{J}_\text{em}$ maps to the fermionic contraint $\langle\sum_{j}{\bf J}^{5 j}\rangle=\sigma\langle\sum_{j} {\bf J}^{j}\rangle$ where $\sigma\rightarrow\infty$.\\
%In summary:
%\begin{center}
%\begin{tabular}{|ccc|}
%\hline
%Physical Effect&Relation&Fermionic Relation\\
%\hline
%No dissipation&${\bf E}=\Lambda\frac{d}{dt}{\bf J}$&$<{\bf J}^5>_F=\Lambda \frac{d}{dt}<{\bf J}>_F$\\
%Dissipation&${\bf J}=\sigma {\bf E}$&$<{\bf J}>_F=\sigma<{\bf J}^5>_F$\\
%Messnier Effect&$B=0$&$<J^5_0>_F=0$\\
%\hline
%\end{tabular}
%\end{center}
{\bf Conclusions.--} In this article we proposed a fermionic tight-binding model which naturally supports the two main ingredients of fermionic superconductivity: Cooper-like pair formation and Meissner effect. In order to prove these effects, we employed functional fermionization to show the equivalence between the effective fermionic theory describing the lattice system (a chiral-invariant Thirring model) and  a double skyrmion model. This model supports skyrmions with bosonic statistics (Cooper-like pairs) in the bi-layer case and  it is formally equivalent to a double Maxwell-BF theory which describes an effective Meissner effect. Moreover, we provide fermionic observables associated to the superfluid phase to detect a signature of such effects in a possible implementation  of the lattice model in a real (or simulated \cite{Sim1,Sim2,Sim3}) quantum system.
A straightforward generalization of our model to the (charged) superconducting case can be obtained once neutral fermions are replaced with charged ones and an external electromagnetic field coupled with them is taken into account. Finally, an open question related to this work concerns the possible existence of Abrikosov-like vortices  and the presence of Majorana states localized at their cores \cite{Read,Nori,Nori2}. We leave the study of these important aspects to future works.

{\bf Acknowledgments.--} 
G. P. thanks Jiannis K. Pachos for comments and useful suggestions and Franco Nori for the hospitality in his group at RIKEN where this work was completed. M.C. thanks Jiannis K. Pachos and Franco Nori for comments and for the financial (and moral!) support in Leeds and Wako-shi where this work was started and completed respectively. 
G. P. is supported by a EPSRC Grant. M.C. is supported by the Canon Foundation in Europe and the RIKEN iTHES program.\\
${}^*$ The authors contributed equally to this article.

\onecolumngrid
\newpage

\section*{\LARGE{Supplemental Material}}
\onecolumngrid
\maketitle
This supplementary material gives the details of the derivations shown in the main text.

\section{Tight-Binding model}
In this section we show that the interacting tight binding model described in the main text is equivalent, at low energy, to a (chiral-invariant) Thirring model.\\
Let us start with the free model: a graphene-like tight binding model with a staggered chemical potential term
\begin{equation}
H^s_0=\pm \left[c \sum_r (a^\dagger_r b_{r+v_1}+a^\dagger_r b_{r+v_2}+a^\dagger_r b_{r})+m c^2\sum_r a^\dagger_r a_{r}-mc^2\sum_r b^\dagger_r b_{r}\right]\;\;,
\end{equation}
where the sign depends on the spin variable, $c$ and $m c^2$ are an energy scales and ${\bf v}_1=(\frac{\sqrt{3}}{2},\frac{3}{2})$ and  ${\bf v}_2=(-\frac{\sqrt{3}}{2},\frac{3}{2})$. The Brillouin zone is defined as $\text{BZ}=\{{\bf p}:p_1 {\bf p}_1+p_2 {\bf p}_2\}$ with $p_1,p_2\in [0,1)$ and ${\bf p}_1=\frac{2}{3\sqrt{3}}(\frac{3}{2},\frac{\sqrt{3}}{2})$, ${\bf p}_2=\frac{2}{3\sqrt{3}}(-\frac{3}{2},\frac{\sqrt{3}}{2})$, so that ${\bf p}=\frac{2}{3\sqrt{3}}(\frac{3}{2}(p_1-p_2),\frac{\sqrt{3}}{2}(p_1+p_2))$ which leads to define ${\bf p}=(p_x,p_y)$, with $p_x=\frac{1}{\sqrt{3}}(p_1-p_2)$ and $p_y=\frac{1}{3}(p_1+p_2)$.
By performing a Fourier transform $a_r=\sum_p e^{2\pi i p r}a_p$, $b_r=\sum_p e^{2\pi i p r}b_p$ we can write
\begin{equation}
H_0=\pm\int d^2p\left(\begin{array}{lll}a^\dagger_p &b^\dagger_p\end{array}\right)\left(\begin{array}{lll}m c^2&f(p)\\f^*(p)&-mc^2\end{array}\right)\left(\begin{array}{l}a_p\\b_p\end{array}\right)\;\;,
\end{equation}
where $f(p)=c(1+e^{-2\pi i p_1}+e^{-2\pi i p_2})$.
%where the Hamiltonian kernel $H$ is
%\begin{equation}
%\bar{H}=\left(\begin{array}{lll}M&f(p)\\f^*(p)&-M\end{array}\right)
%\end{equation}
By solving the equation $f(p)=0$ we find two points in the Brillouin zone for which the kinematic energy is zero. These two Fermi points are: ${\bf P}_+=(\frac{2\pi}{3},\frac{4\pi}{3})$  and ${\bf P}_-=(\frac{4\pi}{3},\frac{2\pi}{3})$. We now want to expand the kinematic term around these two points. In particular we have: $\pp{f}{p_1}_{\big|_+}=-\frac{\sqrt{3}}{2}+\frac{i}{2}$, $\pp{f}{p_2}_{\big|_+}=\frac{\sqrt{3}}{2}+\frac{i}{2}$, $\pp{f}{p_1}_{\big|_-}=\frac{\sqrt{3}}{2}+\frac{i}{2}$, $\pp{f}{p_2}_{\big|_-}=-\frac{\sqrt{3}}{2}+\frac{i}{2}$. By writing ${\bf P}={\bf P}_\pm+k_1{\bf p}_1+k_2{\bf p}_2$ for small $k_1$ and $k_2$ we have, at first order: $f_+=(-\frac{\sqrt{3}}{2}+\frac{i}{2})k_1+(\frac{\sqrt{3}}{2}+\frac{i}{2})k_2=-\frac{3}{2}k_x+\frac{3}{2}i k_y$ and similarly $f_-=(\frac{\sqrt{3}}{2}+\frac{i}{2})k_1+(-\frac{\sqrt{3}}{2}+\frac{i}{2})k_2=\frac{3}{2}k_x+\frac{3}{2}i k_y$ with $k_x=\frac{1}{\sqrt{3}}(k_1-k_2)$ and $k_y=\frac{1}{3}(k_1+k_2)$. This allows us to write the following matrices associated with the Hamiltonian kernels around the two Fermi points as
\begin{equation}
\left\{\begin{array}{lll}
\bar{H}_+&=&-\frac{3}{2}c(\sigma_x k_x+\sigma_y k_y)+m\sigma_z\\
\bar{H}_-&=&-\frac{3}{2}c(\sigma_x k_x-\sigma_y k_y)+m\sigma_z\;\;,
\end{array}\right.
\end{equation}
where $\sigma_{x,y,z}$ are the Pauli matrices and conclude that  the low-energy physics is described by the following Hamiltonian
\begin{equation}
\begin{array}{lll}
H^s_0=\pm\int d^2k\;\Psi_s^{\prime \dagger} \bar{H}\Psi_s^{\prime }\;\;,
\end{array}
\end{equation}
where $\Psi_s^{\prime}=\left(\begin{array}{lllll}a_s^+(k) & b_s^+(k)& a_s^-(k) & b_s^-(k)\end{array}\right)^T$and $\bar{H}=\left(\begin{array}{lll}\bar{H}_+&\hspace{0.1cm}0\\
0&\bar{H}_-\end{array}\right)$.\\
This spinor does not have a form suited for our purposes. In fact, in a later stage, we are interested to add non-trivial interactions among  all the fermions that compose tha spinor (we are interested in chiral interactions involving  a $\gamma_5$ matrix so that we cannot reduce to a 2-spinor). Unfortunately, the spinor presented above involves degrees of freedom evaluated at different Fermi point so that interactions are not readily available. The solution is to  simply mix the spin quantum number with the Fermi point label  to obtain a 4-spinor at each of the Fermi points. This allows us to describe the low-energy limit of the model as
\begin{equation}
H=\int d^2k \; \Psi_+^\dagger \left(\begin{array}{lll}\bar{H}_+&\hspace{0.4cm}0\\0&-\bar{H}_+\end{array}\right)\Psi_++\Psi_-^\dagger \left(\begin{array}{lll}\bar{H}_-&\hspace{0.4cm}0\\0&-\bar{H}_-\end{array}\right)\Psi_-\;\;,
\end{equation}
where $\Psi_\pm=\left(\begin{array}{lllll}a^\uparrow_\pm(k) & b^\uparrow_\pm(k)& a^\downarrow_\pm(k)& b^\downarrow_\pm(k)\end{array}\right)^T$. We can rewrite the previous expression as
\begin{equation}
H=\int d^2 k \left[t\Psi_+^\dagger( k_x\alpha^x+k_y\alpha^y+m\beta)\Psi_+ +t\Psi_-^\dagger  ( -k_x\alpha^x+k_y\alpha^y+m\beta)\Psi_-\right]\;\;,
\end{equation}
where we rescaled $-\frac{3}{2}t\rightarrow t$, $\alpha^{x,y}=\left(\begin{array}{lll}\sigma_{x,y}&\hspace{0.4cm}0\\0&-\sigma_{x,y}\end{array}\right)$ and $\beta=\left(\begin{array}{lll}\sigma_{z}&\hspace{0.3cm}0\\0&-\sigma_{z}\end{array}\right)$. We can now change variables $k_x\rightarrow -k_x$ in the second integral to get
 \begin{equation}
H=\int d^2 k \left[t\Psi_+^\dagger( k_x\alpha^x+k_y\alpha^y+m\beta)\Psi_+ + t\Psi_-^\dagger  ( k_x\alpha^x+k_y\alpha^y+m\beta)\Psi_-\right]\;\;,
\end{equation}
  where we redefined $\Psi_-(k_x,k_y)\rightarrow\Psi_-(-k_x,k_y)$. \\
  We stress that the low energy description of the model is given by a sum of two such action corresponding to the physics around the two Fermi points. Since we do not have any interaction coupling the degrees of freedom around the Fermi points we omit the momentum space label $\pm$ with the agreement that all the following quantities can be evaluated at either Fermi point.
% We now note that the new spinors $\Psi_+$ and $\Psi_-$ contain fermion operators evaluated around the same Fermi point.\\
 We are now ready to introduce the last quantum number. We consider two identical layers (labelled by $j=1,\dots, n$) each one with the structure presented above. This allows us to opportunely introduce inter-layers interaction terms as follows. \\
The full Hamiltonian of the system is
\begin{equation}
H=\sum_j H^j_0+H_I\;\;,
\end{equation}
where 
\begin{equation}
H^j_0=H_0^{j\uparrow}+H_0^{j\downarrow}\;\;,
\end{equation}
with
\begin{equation}
\begin{array}{lll}
H_0^{j\uparrow}&=&\int d^2p \left(\begin{array}{lll}a^{j\uparrow\dagger}_p &b^{j\uparrow\dagger}_p\end{array}\right)\left(\begin{array}{lll}m&f(p)\\f^*(p)&-m\end{array}\right)\left(\begin{array}{l}a^{j\uparrow}_p\\b^{j\uparrow}_p\end{array}\right)\\
H_0^{j\downarrow}&=&-\int d^2p \left(\begin{array}{lll}a^{\downarrow\dagger}_p &b^{j\downarrow\dagger}_p\end{array}\right)\left(\begin{array}{lll}m&f(p)\\f^*(p)&-m\end{array}\right)\left(\begin{array}{l}a^{j\downarrow}_p\\b^{j\downarrow}_p\end{array}\right)\;\;,
\end{array}
\end{equation}
and where we introduced  interaction terms
\begin{equation}
H_I=\frac{g^2}{2}[(\sum_j J^j_\mu)^2+(\sum_j J^{j5}_\mu)^2]\;\;,
\end{equation}
where $J^{j\mu}=\Psi^{\dagger}_j\gamma^\mu\Psi_j$ and $J^{j5\mu}=\Psi^{\dagger}_j\gamma^5\gamma^\mu\Psi_j$ with $\Psi_j=\left(\begin{array}{lllll}a^{j\uparrow} & b^{j\uparrow}& a^{j\downarrow}& b^{j\downarrow}\end{array}\right)^T$. The gamma matrices are defined as $\gamma^0=\beta=\sigma_{z}\otimes \sigma_{z}$, $\gamma^x=\beta\alpha^x=i \sigma_{y}\otimes \mathbb{I}$, $\gamma^y=\beta\alpha^y=i \sigma_{x}\otimes \mathbb{I}$. There is also a further gamma matrix $\gamma^3=i\sigma_{z}\otimes \sigma_{x}$ such that we can define the "fifth" gamma matrix $\gamma^5=i \gamma^0\gamma^1\gamma^2\gamma^3=-\sigma_{z}\otimes \sigma_{y}$. \\
Using the results reported so far, the action describing the low energy of this model is (at each Fermi point)
\begin{equation}
\label{eq:S}
S=S_0+S_I\;\;,
\end{equation}
with
\begin{equation}
\begin{array}{lll}
S_0&=&\int d^3 x\bar{\Psi}_1(i c\slashed{\partial}-mc^2)\Psi_1+\bar{\Psi}_2(i c\slashed{\partial}-mc^2)\Psi_2\\
S_I&=&\int d^3 x\frac{g^2}{2}[(\sum_j J^j_\mu)^2+(\sum_j J^{j5}_\mu)^2]\;\;.
\end{array}
\end{equation}
This shows that, at low energy, the model is described by a chiral-invariant Thirring model. In particular, in the case of a bi-layer ($n=2$), we have that
\begin{equation}
\begin{array}{lll}
H_I&=&3 g^{2}\sum_{j}\left[|(a^{j\uparrow}b^{j\uparrow}+a^{j\downarrow}b^{j\downarrow})|^{2}+|(a^{j\uparrow}b^{j\downarrow}-a^{j\downarrow}b^{j\uparrow})|^{2}+|(a^{1\uparrow}a^{2\downarrow}+a^{1\downarrow}a^{2\uparrow})|^{2}+|(a^{1\uparrow}a^{2\uparrow}-a^{1\downarrow}a^{2\downarrow})|^{2}\right. \\
&&\left.+|(b^{1\uparrow}b^{2\downarrow}+b^{1\downarrow}b^{2\uparrow})|^{2}+|(b^{1\uparrow}b^{2\uparrow}-b^{1\downarrow}b^{2\downarrow})|^{2}+|(a^{1\uparrow}b^{2\downarrow}+b^{1\uparrow}a^{2\downarrow}-a^{1\downarrow}b^{2\uparrow}-b^{1\downarrow}a^{2\uparrow})|^{2}+ \right. \\ 
&&\left.|(a^{1\uparrow}b^{2\uparrow}+a^{1\downarrow}b^{2\downarrow}+b^{1\uparrow}a^{2\uparrow}+b^{1\downarrow}a^{2\downarrow})|^{2}+(a^{2\downarrow}b^{1\uparrow}-a^{2\uparrow}b^{1\downarrow})^{\dagger}(a^{1\downarrow}b^{2\uparrow}-a^{1\uparrow}b^{2\downarrow})+\right. \\ 
&&\left.(b^{2\uparrow}a^{1\uparrow}+b^{2\downarrow}a^{1\downarrow})^{\dagger}(b^{1\uparrow}a^{2\uparrow}+b^{1\downarrow}a^{2\downarrow})+\text{h.c.}\right]\;\;.
\end{array}
\end{equation}

%\newpage
\section{Double skyrmion model and functional fermionization}
In this section, we show that a double $O(3)$-Hopf non-linear sigma model (being equivalent to a double CP-CS theory) is mapped in a chiral-invariant Thirring model by generalizing the fermionization techniques introduced in \cite{Mavromatos2,Huerta}. Moreover, we derive \emph{fermionization rules} which map observables of the fermionic model to those ones of the bosonic (double CP-CS) model.
\subsection{Double skyrmion model and double CP-CS theory}
The double skyrmion model is defined by the following partition function
\begin{equation}
\label{eq:O3Hopf}
Z_{(\text{O}(3)-\text{H})^2}=\int \mathcal{D}{\bf m}^+\mathcal{D}{\bf m}^-  \exp\left\{i\int d^3 x\left[\frac{1}{2g_0^2}(\partial_\mu{\bf m}^+\partial^\mu {\bf m}^+)+n\pi H^++\frac{1}{2g_0^2}(\partial_\mu{\bf m}^-\partial^\mu {\bf m}^-)-n\pi H^-\right]\right\}\;\;,
\end{equation}
with the constraint ${\bf m}^2=1$ and $H$ is the Hopf invariant defined as \cite{Abanov1}
\begin{equation}
H^\pm=\frac{\epsilon^{\mu\nu\lambda}}{24\pi^2}\int d^3 x~\text{tr}\left[(U_\pm^{-1}\partial_\mu U_\pm)(U_\pm^{-1}\partial_\nu U_\pm)(U_\pm^{-1}\partial_\lambda U_\pm)\right]\;\;,
\end{equation}
where  $\sum_i{\bf m}^\pm_i {\bf \sigma}_i=U_\pm^{-1}\sigma_3 U_\pm$.\\
On the other hand, a double CP-CS model is defined by the partition function
\begin{equation}
\label{eq:2CPCS}
\begin{array}{lll}
Z_{\text{(CP-CS)}^2}&=&\int \mathcal{D}A^+\mathcal{D}A^-\mathcal{D}z^+ \mathcal{D}z^-\exp\{i \int d^3 x\left[  \frac{n}{4\pi}\epsilon^{\lambda\mu\nu}A^+_\lambda\partial_\mu A^+_\nu- \frac{n }{4\pi}\epsilon^{\lambda\mu\nu}A^-_\lambda\partial_\mu A^-_\nu\right.\\
&&\left.+\frac{1}{{g^2}}\int d^3 x |(i\partial_\mu-A^+_\mu)z^+|^2+\frac{1}{{g^2}}\int d^3 x |(i\partial_\mu-A^-_\mu)z^-|^2\right]\}\;\;,
\end{array}
\end{equation}
where  $z^\pm=\left(\begin{array}{l}z^\pm_1~~z^\pm_2\end{array}\right)^T$ with the fields $z^\pm_1,z^\pm_2\in\mathbb{C}$ such that $|z^\pm|^2=z^\pm_1z^{\pm*}_1+z^\pm_2z^{\pm*}_2=1$.\\
The low-energy equivalence of these two  models comes from using a saddle point approximation to integrate the fields $A^+$ and $A^-$. This results in the following identities \cite{Fradkin1}
\begin{equation}
\label{eq_identities}
\left\{\begin{array}{clc}
\frac{1}{g_0^2} |(\partial_\mu-A^\pm_\mu)z^\pm|^2&=& \frac{1}{2g_0^2}(\partial_\mu{\bf m}^\pm\partial^\mu {\bf m}^\pm)\\
A^\pm_\mu&=&-\frac{i}{2}z^{\pm*}\partial_\mu z^\pm\\
{\bf m}^\pm&=&z^{\pm*}_\alpha{\bf \sigma}_{\alpha\beta}z^{\pm}_\beta\\
\frac{1}{4\pi}\int d^3 x\epsilon^{\lambda\mu\nu}A^\pm_\lambda\partial_\mu A^\pm_\nu&=&\pi H^\pm\;\;.
\end{array}\right.
\end{equation}
%Now, from the double Chern-Simons theory we find that the spin associated with the theory is
%\begin{equation}
%s=\frac{Q^2}{2 N}
%\end{equation}
%where $Q$ is the ``electric'' charge in units of $e$ and $n=2$ corresponds to the two fermionic species present in the model. On the other hand the spin associated with the double skyrmion model is
%By consistency, we find
%\begin{equation}
%Q=N Q_T
%\end{equation}
%with $n=2$.
The spin associated with the double skyrmion model is given by \cite{Karabali}
\begin{equation}
S =n \frac{ (Q^\pm_T)^2}{2}\;,
\end{equation}
where $Q^\pm_T=\int d^2 x J^{0\pm}_S$ is the topological charge with $J^{0\pm}_S$  the 0th components of the two skyrmion currents $J_{S}^{\mu\pm}=\frac{1}{8 \pi}\epsilon^{\mu\nu\lambda}\epsilon^{abc}m^{\pm}_a\partial_\nu m^{\pm}_b \partial_\lambda m^{\pm}_c=\frac{1}{2\pi}\epsilon^{\mu\nu\lambda}\partial_{\nu}A_{\lambda}$ \cite{Fradkin1}. Due to the different sign in front of the Hopf terms, this theory describes independent skyrmions and anti-skyrmions which have opposite values of the topological charges $Q^+_T=-Q^-_T$ which assume only integer values. Skyrmions have fermionic or bosonic statistics depending on the value of $n$. For odd $n$, skyrmions are fermions so that the emergence of fermionic superfluidity must be related to a condensation into bosons. Insted, when $n$ is even, skyrmions behave like bosons and already simulate the Cooper-like pairs properties.
\subsection{Functional fermionization}
\label{section:Fermionization}
Following \cite{Mavromatos2}, we now want to show that a theory described by a double CP-Chern-Simons is equivalent to a chiral-invariant Thirring model.  Our starting point is the partition function
\begin{equation}
\label{eq:CCPP-CCSS}
\begin{array}{lll}
Z_{\text{(CP-CS)}^2}&=&\int \mathcal{D}A^+\mathcal{D}A^-\mathcal{D}z^+ \mathcal{D}z^-\exp\{i \int d^3 x\left[  \frac{n}{4\pi}\epsilon^{\lambda\mu\nu}A^+_\lambda\partial_\mu A^+_\nu- \frac{n }{4\pi}\epsilon^{\lambda\mu\nu}A^-_\lambda\partial_\mu A^-_\nu\right.\\
&&\left.+\frac{1}{{g^2}}\int d^3 x |(i\partial_\mu-A^+_\mu)z^+|^2+\frac{1}{{g^2}}\int d^3 x |(i\partial_\mu-A^-_\mu)z^-|^2\right]\}\;\;,
\end{array}
\end{equation}
where  $z^\pm=\left(\begin{array}{l}z^\pm_1\\z^\pm_2\end{array}\right)$ where $z^\pm_1,z^\pm_2\in\mathbb{C}$ such that $|z^\pm|^2=z^\pm_1z^{\pm*}_1+z^\pm_2z^{\pm*}_2=1$.
We now perform the following change of variables
\begin{equation}
\label{eq:changeOfVariablesCPCS}
\left\{\begin{array}{lll}
A_\mu^+&=&A_\mu+B_\mu\\
A_\mu^-&=&A_\mu-B_\mu\;\;,
\end{array}\right.
\end{equation}
so that, after an integration by parts on a manifold without boundary we get $\epsilon^{\mu\nu\lambda}(A_\mu^+\partial_\nu A_\lambda^+-A_\mu^-\partial_\nu A_\lambda^-)=4 \epsilon^{\mu\nu\lambda}B_\mu\partial_\nu A_\lambda$ and an equivalent partition function
\begin{equation}
\begin{array}{lll}
Z_{\text{(CP)}^2- BF}&=&\int \mathcal{D}A\mathcal{D}B\mathcal{D}z^+ \mathcal{D}z^-\exp\left\{i \int d^3 x\left[  \frac{n}{\pi}\epsilon^{\mu\nu\lambda}B_\mu\partial_\nu A_\lambda\right.\right.\\
&&\left.\left.+\frac{1}{{g^2}} |(i\partial_\mu-(A_\mu+B_\mu))z^+|^2+\frac{1}{{g^2}} |(i\partial_\mu-(A_\mu-B_\mu))z^-|^2\right]\right\}\;\;.
\end{array}
\end{equation}
Now, we can introduce $n$ species of fermions in order to ``linearize'' the BF term \cite{Mavromatos1}
\begin{equation}
\label{eq:linearization}
e^{i\frac{n}{\pi}\int d^3 x \epsilon^{\mu\nu\lambda}B_\mu \partial_\nu A_\lambda}=\int \mathcal{D}\chi\mathcal{D}\bar{\chi}e^{i\sum_j^n(\bar{\chi}_j (i\slashed{\partial}-m)\chi_j-\sqrt{2}A_\mu J_\chi^{j\mu}-\sqrt{2}B_\mu J_\chi^{5j\mu})}\;\;,
\end{equation}
where we used the identity $e^{i\frac{n}{2\pi}\int d^3 x \epsilon^{\mu\nu\lambda}B_\mu \partial_\nu A_\lambda}=\int \mathcal{D}\chi\mathcal{D}\bar{\chi}e^{i\sum_j^n(\bar{\chi}_j (i\slashed{\partial}-m)\chi_j-A_\mu J_\chi^{j\mu}-B_\mu J_\chi^{5j\mu})}$ and where $J_\chi^{j\mu}=\bar{\chi}_j\gamma^\mu\chi_j$ and $J_\chi^{5j\mu}=\bar{\chi}_j\gamma^5\gamma^\mu\chi_j$. In this way the partition function becomes
\begin{equation}
\label{eq:endEasyRules}
\begin{array}{lll}
Z&=&\int \mathcal{D}A\mathcal{D}B\mathcal{D}z^+ \mathcal{D}z^-\mathcal{D}\chi\mathcal{D}\bar{\chi}\exp\{i \int d^3 x\left[ \sum_j(\bar{\chi}_j (i\slashed{\partial}-m)\chi_j-\sqrt{2}A_\mu J_\chi^{j\mu}-\sqrt{2}B_\mu J_\chi^{5j\mu})\right.\\
&&\left.+\frac{1}{{g^2}}|(i\partial_\mu-(A_\mu+B_\mu))z^+|^2+\frac{1}{{g^2}} |(i\partial_\mu-(A_\mu-B_\mu))z^-|^2\right]\}\;\;.
\end{array}
\end{equation}
We now consider the following change of variables and notation
\begin{equation}
z^\pm\rightarrow \tilde{Z}^\pm\equiv\left(\begin{array}{ccc}
z^\pm_1&-z_2^{\pm*}\\
z^\pm_2&z_1^{\pm*}
\end{array}\right)
\equiv e^{i\xi^\pm_j\sigma^j}\;\;,
\end{equation}
%\begin{equation}
%\begin{array}{lll}
%z^\pm&=&e^{i\xi^\pm_j\sigma^j}\\
%A_\mu&\rightarrow&A_\mu\sigma^3\\
%B_\mu&\rightarrow&B_\mu\sigma^3\\
%\end{array}
%\end{equation}
which allows us to write (omitting the $\pm$ labels) the CP terms as
\begin{equation}
[(i\partial_\mu-(A_\mu+B_\mu)]z|^2\rightarrow\frac{1}{2}\text{Tr}|[\partial_\mu-(A_\mu+B_\mu)\sigma_3]\tilde{Z}|^2\;\;,
\end{equation}
where
\begin{equation}
|[i\partial_\mu-(A_\mu+B_\mu)\sigma_3]\tilde{Z}|^2\equiv \tilde{Z}^\dagger [-i\cev{\partial}_\mu-(A_\mu+B_\mu)\sigma_3]\cdot [i\vec{\partial}_\mu-(A_\mu+B_\mu)\sigma_3]\tilde{Z}\;\;.
\end{equation}
The $z$-dependent terms can now be rewritten as
\begin{equation}
\begin{array}{lll}
\frac{1}{2}\text{Tr}|[i\partial_\mu-(A_\mu+B_\mu)\sigma_3] Z|^2&=&\frac{1}{2}\text{Tr}|[-\partial_\mu\xi_j\sigma^j-(A_\mu+B_\mu)\sigma_3]\tilde{Z}|^2\\
&=&\frac{1}{2}\text{Tr}(\partial_\mu\xi_j\sigma^j)(\partial^\mu\xi_j\sigma^j)+(A_\mu+B_\mu)^2+\partial_\mu\xi_j(A^\mu+B^\mu)(\sigma^j\sigma_3+\sigma_3 \sigma^j)\\
&=&\frac{1}{2}\text{Tr}[(\partial_\mu\xi_j)^2+(A_\mu+B_\mu)^2+J^{\xi i}_\mu (A^\mu+B^\mu)(\sigma^j\sigma_3+\sigma_3 \sigma^j)]\\
&=&[(\partial_\mu\xi_j)^2+(A_\mu+B_\mu)^2+J^{\xi 3}_\mu (A^\mu+B^\mu)]\;\;,
\end{array}
\end{equation}
where $J^{\xi i}_\mu=2\partial_\mu \xi^i$.
With this transformation the action becomes
\begin{equation}
\begin{array}{lll}
Z&=&\int \mathcal{D}A\mathcal{D}B\mathcal{D}\xi^+ \mathcal{D}\xi^-\mathcal{D}\chi\mathcal{D}\bar{\chi}\exp\left\{i \int d^3 x\left[ \sum_j(\bar{\chi}_j (i\slashed{\partial}-m)\chi_j-\sqrt{2}A_\mu J_\chi^{j\mu}-\sqrt{2}B_\mu J_\chi^{5j\mu})\right.\right.\\
&&\left.\left.+\frac{1}{{g^2}} \left\{(\partial_\mu \xi^+_j)^2+(A_\mu+B_\mu)^2+ J^{{\xi^+} 3}_\mu (A^\mu+B^\mu)\right.\right.+\left.\left.\frac{1}{{g^2}}(\partial_\mu \xi^-_j)^2+(A_\mu-B_\mu)^2+ J^{{\xi^-} 3}_\mu (A^\mu-B^\mu)\right\}\right]\right\}\\
&=&\int \mathcal{D}A\mathcal{D}B\mathcal{D}\xi^+ \mathcal{D}\xi^-\mathcal{D}\chi\mathcal{D}\bar{\chi}\exp\left\{i \int d^3 x\left[ \sum_j(\bar{\chi}_j (i\slashed{\partial}-m)\chi_j-\sqrt{2}A_\mu J_\chi^{j\mu}-\sqrt{2} B_\mu J_\chi^{5j\mu})\right.\right.\\
&&+\frac{1}{{g^2}}\left\{ (\partial_\mu \xi^+_j)^2+(\partial_\mu \xi^-_j)^2+A^2_\mu+B_\mu^2+\left.\left.A^\mu(J^{\xi^+ 3}_\mu+J^{\xi^- 3}_\mu)+B^\mu(J^{\xi^+ 3}_\mu-J^{\xi^- 3}_\mu)\right\}\right]\right\}\;\;.
\end{array}
\end{equation}
We now change variables as
\begin{equation}
\left\{\begin{array}{lll}
\xi^A_j&=&\xi^+_j+\xi^-_j\\
\xi^B_j&=&\xi^+_j-\xi^-_j\;\;.
\end{array}\right.
\end{equation}
so that
\begin{equation}
\begin{array}{lll}
Z
&=&\int \mathcal{D}A\mathcal{D}B\mathcal{D}\xi^+ \mathcal{D}\xi^-\mathcal{D}\chi\mathcal{D}\bar{\chi}\exp\left\{i \int d^3 x\left[ \sum_j(\bar{\chi}_j (i\slashed{\partial}-m)\chi_j-\sqrt{2}A_\mu J_\chi^{j\mu}-\sqrt{2}B_\mu J_\chi^{5j\mu})\right.\right.\\
&&+\frac{1}{{g^2}} \left\{\frac{1}{2}(\partial_\mu \xi^A_j)^2+\frac{1}{2}(\partial_\mu \xi^B_j)^2+A^2_\mu+B_\mu^2+\left.\left. A^\mu J^{\xi^A 3}_\mu+B^\mu J^{\xi^B 3}_\mu\right\}\right]\right\}\\
&=&\int \mathcal{D}A\mathcal{D}B\mathcal{D}\xi^+ \mathcal{D}\xi^-\mathcal{D}\chi\mathcal{D}\bar{\chi}\exp\left\{i \int d^3 x\left[ \sum_j(\bar{\chi}_j (\slashed{\partial}-m)\chi_j\right.+\frac{1}{{g^2}}\left\{\frac{1}{2}(\partial_\mu \xi^A_j)^2+\frac{1}{2}(\partial_\mu \xi^B_j)^2\right.\right.\\
&&-
\left.\left.\frac{1}{{g^2}}(A_\mu^2+B_\mu^2)+A^\mu (-\sqrt{2}\sum_j J_\mu^ {j\chi}+ \frac{1 }{{g^2}}J^{\xi^A 3}_\mu)\right.\left.+B^\mu (-\sqrt{2}\sum_j J_\mu^{5 j\chi}+ \frac{1 }{{g^2}}J^{\xi^B 3}_\mu)\right]\right\}\;\;,
\end{array}
\end{equation}
where
\begin{equation}
\left\{\begin{array}{lllll}
J^{\xi^A 3}_\mu&=&J^{\xi^+ 3}_\mu+J^{\xi^- 3}_\mu&=&2\partial_\mu\xi_j^A\\
J^{\xi^B 3}_\mu&=&J^{\xi^+ 3}_\mu-J^{\xi^- 3}_\mu&=&2\partial_\mu\xi_j^B\;\;.
\end{array}\right.
\end{equation} 
Thanks to the general gaussian integral identity $\int\mathcal{D} A e^{-i\int d^3 x (-\frac{1}{2g^2}A_\mu A^\mu+J_\mu A^\mu)}=e^{-i\int d^3 x\frac{g^2}{2}J_\mu J^\mu}$ we get
\begin{equation}
\int\mathcal{D} A e^{i\int d^3 x (\frac{1}{{g^2}}A_\mu A^\mu+\tilde{J^A}_\mu A^\mu)}=e^{-i\int d^3 x\frac{{g^2}}{4}\tilde{J}^A_\mu \tilde{J}^{A\mu}}\;\;,
\end{equation}
 and
\begin{equation}
\int\mathcal{D} B e^{i\int d^3 x (\frac{1}{{g^2}}B_\mu B^\mu+\tilde{J^B}_\mu B^\mu)}=e^{-i\int d^3 x\frac{{g^2}}{4}\tilde{J}^B_\mu \tilde{J}^{B\mu}}\;\;.
\end{equation}
By identifying  $\tilde{J}^A_\mu=-\sqrt{2}\sum_j J_\mu^ {j\chi}+ \frac{1 }{{g^2}}J^{\xi^A 3}_\mu$ and  $\tilde{J}^B_\mu=-\sqrt{2}\sum_j J_\mu^{5 j\chi}+ \frac{1 }{{g^2}}J^{\xi^B 3}_\mu$, we can integrate over $A$ and $B$ to get
\begin{equation}
\begin{array}{lll}
Z
&=&\int \mathcal{D}\xi^+ \mathcal{D}\xi^-\mathcal{D}\chi\mathcal{D}\bar{\chi}\exp\{i \int d^3 x\left[ \sum_j(\bar{\chi}_j (i\slashed{\partial}-m)\chi_j\right.+\frac{1}{{g^2}}\left\{ \frac{1}{2}(\partial_\mu \xi^A_j)^2+\frac{1}{2}(\partial_\mu \xi^B_j)^2\right\}\\
&&\left.-\frac{{g^2}}{4}(\frac{1}{{g^2}}J_\mu^{3A}-\sqrt{2}\sum_j J^{j\chi}_\mu))^2-\frac{{g^2}}{4}((\frac{1}{{g^2}}J_\mu^{3B}-\sqrt{2}\sum_j J^{j5\chi}_\mu))^2\right]\\
&=&\int \mathcal{D}\xi^+ \mathcal{D}\xi^-\mathcal{D}\chi\mathcal{D}\bar{\chi}\exp\{i \int d^3 x\left[ \sum_j(\bar{\chi}_j (i\slashed{\partial}-m)\chi_j\right.+\frac{1}{{g^2}}\left\{ \frac{1}{2}(\partial_\mu \xi^A_j)^2+\frac{1}{2}(\partial_\mu \xi^B_j)^2\right\}\\
&&\frac{{g^2}}{2}\left((\sum_j J^{\chi j}_\mu)^2+(\sum_j J^{5\chi j}_\mu)^2\right)-
(\frac{1}{4{g^2}})\left(J^{3A}_\mu+J^{3B}_\mu\right)+\frac{1}{\sqrt{2}}\left(J_\mu^{3 A}(\sum_j J_\mu^{j\chi})+J_\mu^{3 B}(\sum_j J_\mu^{5 j\chi})\right)\;\;.
\end{array}
\end{equation}
We now change spinor variables $\chi=\Theta\Psi$, with $\Theta=e^{i \sqrt{2} (\xi^{3A}+ \gamma^5 \xi^{3B})}$  so that the kinematic part $\bar{\chi}_j (i\slashed{\partial}-m)\chi_j$ of the fermionic action gives us a piece
\begin{equation}
\begin{array}{l}
-\sqrt{2}(\bar{\Psi}_j  \gamma^\mu\partial_\mu \xi^{3A}\Psi_j+\bar{\Psi}_j \gamma^\mu\partial_\mu \xi^{3A}\gamma^5\Psi_j)=-\frac{1}{\sqrt{2}}\left(J_\mu^{3A}(\sum_jJ^{j\Psi}_\mu)+J_\mu^{3B}(\sum J^{5j\Psi}_\mu)\right)\;\;,
\end{array}
\end{equation}
which simplifies the expression for the partition function to
\begin{equation}
\begin{array}{lll}
Z&=&\int \mathcal{D}\xi^+ \mathcal{D}\xi^-\mathcal{D}\Psi\mathcal{D}\bar{\Psi}\exp\left\{i \int d^3 x\left[ \sum_j(\bar{\Psi}_j (i\slashed{\partial}-m)\Psi_j\right.+\frac{1}{{g^2}}\int d^3 x \frac{1}{2}(\partial_\mu \xi^A_j)^2+\frac{1}{2}(\partial_\mu \xi^B_j)^2\right.\\
&&\left.\left.+\frac{{g^2}}{2}(\sum_j J^{\Psi j}_\mu)^2+\frac{{g^2}}{2}(\sum_j J^{5\Psi j}_\mu)^2-
(\frac{1}{4{g^2}})^2J^{3A}_\mu-(\frac{1}{4{g^2}})^2J^{3B}_\mu\right]\right\}\;\;,
\end{array}
\end{equation}
where we used the fact that
\begin{equation}
\label{eq:equalCurrents}
\left\{\begin{array}{lll}
J^{j\chi}_\mu &=&J_{\mu}^{j\Psi}\\
J^{j5\chi}_\mu &=&J_{\mu}^{j5\Psi}\;\;.
\end{array}\right.
\end{equation}
We then see that the fields $\xi$ do  not interact with the fermion so that we can integrate them out to get
\begin{equation}
\label{eq:ThirringAfterFermionization}
\begin{array}{lll}
Z_F=\int \mathcal{D}\Psi\mathcal{D}\bar{\Psi}\exp\{i \int d^3 x\left[ \sum_j(\bar{\Psi}_j (i\slashed{\partial}-m)\Psi_j\right. \left.+\frac{{g^2}}{2}(\sum_j J^{\Psi j}_\mu)^2+\frac{{g^2}}{2}(\sum_j J^{5\Psi j}_\mu)^2\right]\;\;,
\end{array}
\end{equation}
which is in fact the original chiral-invariant Thirring model we started from.
\subsection{Fermionization rules}
In this subsection we map observables for the Thirring model and observables for the bosonic double CP-CS theory.\\
To this end, we begin by introducing external fields in the fermionic theory of Eq. (\ref{eq:ThirringAfterFermionization}) via a minimal coupling
\begin{equation}
\label{eq:ThirringAfterFermionization2}
\begin{array}{lll}
Z_F(J^\Psi,J^{5\Psi})&=&\int \mathcal{D}\Psi\mathcal{D}\bar{\Psi}\exp\{i \int d^3 x\left[ \sum_j(\bar{\Psi}_j (i\slashed{\partial}-m)\Psi_j\right.\frac{{g^2}}{2}(\sum_j J^{\Psi j}_\mu)^2+\frac{{g^2}}{2}(\sum_j J^{5\Psi j}_\mu)^2\\
&&\left.+\sum_j J^{ j}_\mu A^{\mu}_\text{ext}+\sum_j J^{5 j}_\mu B^{\mu}_\text{ext}\right]\;\;,
\end{array}
\end{equation}
where we omitted the label $\Psi$ in the currents in light of the identities in Eq. (\ref{eq:equalCurrents}).\\
We now notice that all the steps done in the fermionization process described in the previous subsection can be reversed. We then replace the Thirring action in Eq. (\ref{eq:ThirringAfterFermionization}) with Eq.(\ref{eq:ThirringAfterFermionization2}) and follow all the fermionization steps back. The newly introduced terms depending on the external fields can be carried over until Eq. (\ref{eq:endEasyRules}) by simply replacing
\begin{equation}
i\slashed{\partial}-m\rightarrow i\slashed{\partial}-m+\slashed{A}_{\text{ext}}+\gamma^5 \slashed{B}_{\text{ext}}\;\;.
\end{equation}
We can then change variables to
\begin{equation}
\left\{\begin{array}{lll}
\bar{A}^\mu&=&A^\mu-\frac{1}{\sqrt{2}}A^\mu_\text{ext}\\
\bar{B}^\mu&=&B^\mu-\frac{1}{\sqrt{2}}B^\mu_\text{ext}\;\;.
\end{array}\right.
\end{equation}
before integrating out the fermions. This allows us to use Eq. (\ref{eq:linearization}) as it is and to get, in place of Eq. (\ref{eq:CCPP-CCSS})
\begin{equation}
\begin{array}{lll}
Z_{\text{CP}^2-BF}(J^\Psi,J^{5\Psi})&=&\int \mathcal{D}\bar{A}\mathcal{D}\bar{B}\mathcal{D}z^+ \mathcal{D}z^-\exp\{i \int d^3 x\left[  \frac{n}{\pi}\epsilon_{\mu\nu\lambda}\bar{B}^\mu\partial^\nu\bar{A}^\lambda\right.\\
&&\left.+\frac{1}{{g^2}} |(i\partial_\mu-(\bar{A}_\mu+\bar{B}_\mu+\frac{1}{\sqrt{2}}A_\mu^\text{ext}+\frac{1}{\sqrt{2}}B_\mu^\text{ext})z^+|^2\right.\\
&&\left.+\frac{1}{{g^2}} |(i\partial_\mu-(\bar{A}_\mu-\bar{B}_\mu+\frac{1}{\sqrt{2}}A_\mu^\text{ext}-\frac{1}{\sqrt{2}}B_\mu^\text{ext}))z^-|^2\right]\}\\
&=&\int \mathcal{D}A\mathcal{D}B\mathcal{D}z^+ \mathcal{D}z^-\exp\{i \int d^3 x\left[  \frac{n}{\pi}\epsilon_{\mu\nu\lambda}(B^\mu-\frac{1}{\sqrt{2}}B^\mu_\text{ext})\partial^\nu(A^\lambda-\frac{1}{\sqrt{2}}A^\lambda_\text{ext})\right.\\
&&\left.+\frac{1}{{g^2}} |(i\partial_\mu-(A_\mu+B_\mu))z^+|^2+\frac{1}{{g^2}} |(i\partial_\mu-(A_\mu-B_\mu))z^-|^2\right]\}\\
&=&\int \mathcal{D}A\mathcal{D}B\mathcal{D}z^+ \mathcal{D}z^-\exp\{i \int d^3 x\left[  \frac{n}{\pi}\epsilon_{\mu\nu\lambda}B^\mu\partial^\nu A^\lambda\right.\\
&&\left.+\frac{1}{{g^2}} |(i\partial_\mu-(A_\mu+B_\mu))z^+|^2+\frac{1}{{g^2}} |(i\partial_\mu-(A_\mu-B_\mu))z^-|^2\right\}\\
&&\left.+\frac{n}{\sqrt{2}\pi}\epsilon_{\mu\nu\lambda}(-B^\mu_\text{ext}\partial^\nu A^\lambda+A^\mu_\text{ext}\partial_\nu B_\lambda)+\frac{n}{2\pi}\epsilon_{\mu\nu\lambda}B^\mu_\text{ext}\partial^\nu A^\lambda_\text{ext}\right]\;\;.
\end{array}
\end{equation}
This immediately allows us to prove the following fermionization rules
\begin{equation}
\label{eq:fermRules}
\left\{\begin{array}{lll}
\sum_j J_\mu^j&\leftrightarrow&\frac{n}{\sqrt{2}\pi}\epsilon_{\mu\nu\lambda}\partial^\nu B^\lambda\\
\sum_j J_\mu^{5 j}&\leftrightarrow& - \frac{n}{\sqrt{2}\pi}\epsilon_{\mu\nu\lambda}\partial^\nu A^\lambda\;\;,
\end{array}\right.
\end{equation}
which hold in the following sense
\begin{equation}
\label{eq:fermionizationRules1}
\left\{\begin{array}{lll}
\sum_j\langle J_\mu^j\rangle_F&=&\frac{n}{\sqrt{2}\pi}\epsilon_{\mu\nu\lambda}\langle \partial^\nu B^\lambda\rangle_{\text{CP}^2-BF}\\
\sum_j\langle J_\mu^{5 j}\rangle_F&=&-\frac{n}{\sqrt{2}\pi}\epsilon_{\mu\nu\lambda}\langle \partial^\nu A^\lambda\rangle_{\text{CP}^2-BF}\;\;,
\end{array}\right.
\end{equation}
where the expectation values $\langle\rangle_F$ and $\langle\rangle_{\text{CP}^2-BF}$ are calculated with respect to the ground state of the  fermionic and bosonic theory respectively. The proof for this is a simple consequence of the duality $Z_F(J^\Psi,J^{5\Psi})=Z_{\text{CP}^2-BF}(J^\Psi,J^{5\Psi})$. In fact
\begin{equation}
\begin{array}{lllllll}
\langle J_\mu\rangle_F&=&\frac{\delta Z_F}{\delta A^\mu_{ext}}_{{\Big|}_{(A,B)_\text{ext}=0}}&=&\frac{\delta Z_{\text{CP}^2-BF}}{\delta A^\mu_{ext}}_{{\Big|}_{(A,B)_\text{ext}=0}}&=&\frac{n}{\sqrt{2}\pi}\epsilon_{\mu\nu\lambda}\langle \partial^\nu B^\lambda\rangle_{\text{CP}^2-BF}\\
\langle J_\mu^5\rangle_F&=&\frac{\delta Z_F}{\delta B^\mu_{ext}}_{{\Big|}_{(A,B)_\text{ext}=0}}&=&\frac{\delta Z_{\text{CP}^2-BF}}{\delta B^\mu_{ext}}_{{\Big|}_{(A,B)_\text{ext}=0}}&=&-\frac{n}{\sqrt{2}\pi}\epsilon_{\mu\nu\lambda}\langle \partial^\nu A^\lambda\rangle_{\text{CP}^2-BF}\;\;.
\end{array}
\end{equation}
%\newpage
\section{Superfluidity physics}
In this section we show that the low energy physics of the model is described by a London action which imply a Meissner effect for the effective magnetic field and dissipationless currents. We then associate effective superfluidity effects to fermionic identities between observables for the original tight binding model.
\subsection{London Action}
In this subsection we describe the low-energy physics of the model with a London action.\\
In \cite{Mavromatos1}, it is proven that (at low energy) a Maxwell theory is equivalent to a $CP$ model, namely
\begin{equation}
\int\mathcal{D}A\mathcal{D}z\mathcal{D}z^\dagger\delta(z^\dagger z-1)e^{\frac{i}{g_0^2}\int d^3 x |(i\partial_\mu-A_\mu)z|^2}=\int \mathcal{D}A e^{-\frac{i}{4e^2}\int d^3 x F(A)_{\mu\nu}F(A)^{\mu\nu}}\;\;,
\end{equation}
where $z=(z_1,z_2)$ with $z_1,z_2\in\mathbb{C}$ and $|z_1|^2+|z_2|^2=1$ and $e^2=24\pi |M|$ where $M$ is given by the consistency condition
\begin{equation}
1=i g_0^2\int \frac{d^3 k}{(2\pi)^3}\frac{1}{k^2-M^2}\;\;,
\end{equation}
which renormalizes the coupling strength in relation to a momentum cut-off $|k|=\Lambda$ as
\begin{equation}
1=ig_0^2\int_0^\Lambda \frac{\sin{\theta}d^3 k}{(2\pi)^3}\frac{k^2}{k^2-M^2}=i g_0^2\frac{4\pi}{(2\pi)^3}\int_0^\Lambda\frac{k^2}{k^2-M^2}=i g_0^2\frac{4\pi}{(2\pi)^3}(\Lambda-m\arctan{\frac{\Lambda}{M}})\;\;,
\end{equation}
which, for $\Lambda=\sqrt[3]{\frac{3}{2}}~s ~M$ with $s\ll 1$ (low kinetic energy limit) allows us to write
\begin{equation}
\begin{array}{lll}
1&=&i g_0^2\frac{4\pi}{(2\pi)^3}(\Lambda-M(\frac{\Lambda}{M}+\frac{1}{3}\frac{\Lambda^3}{M^3}M))\;\;,
\end{array}
\end{equation}
which leads to
\begin{equation}
|M|=\frac{(2\pi)^2}{s g_0^2}\;\;,
\end{equation}
or, equivalently
\begin{equation}
\label{eq:ee}
e^2=24 \pi \frac{(2\pi)^2}{s g_0^2}\;\;.
\end{equation}
 By using this mapping we can map our double (CP-CS) to a double (CS-Maxwell) theory
\begin{equation}
\begin{array}{lll}
S_{\text{M-CS}^2}=\int d^3 x\left[  \frac{n}{4\pi}\epsilon^{\lambda\mu\nu}A^+_\lambda\partial_\mu A^+_\nu- \frac{n }{4\pi}\epsilon^{\lambda\mu\nu}A^-_\lambda\partial_\mu A^-_\nu\right.
\left.-\frac{1}{4 e^2} F_{\mu\nu}(A^{+})F^{\mu\nu}(A^{+})-\frac{1}{4 e^2} F_{\mu\nu}(A^{-})F^{\mu\nu}(A^{-})\right]\;.
\end{array}
\end{equation}
By defining new fields $A$ and $B$ as $A^+_\mu=A_\mu+B_\mu$ and $A^-_\mu=A_\mu-B_\mu$, we get the dual theory with action
\begin{equation}
\begin{array}{lll}
S_{\text{M}^2-BF}&=&\int d^3 x\left[  \frac{n}{\pi}\epsilon^{\lambda\mu\nu}B_\lambda\partial_\mu A^+_\nu-\frac{1}{4 e^2}F_{\mu\nu}(A)F^{\mu\nu}(A)-\frac{1}{4 e^2}F_{\mu\nu}(B)F^{\mu\nu}(B)\right]\;\;,
\end{array}
\end{equation}
so that our theory is defined by
\begin{equation}
\label{eq:bosonicTheory}
\begin{array}{lll}
Z_{\text{M}^2-BF}&=&\int \mathcal{D}A\mathcal{D}B e^{i\int d^3 x\left[  \frac{n}{\pi}\epsilon^{\lambda\mu\nu}B_\lambda\partial_\mu A_\nu-\frac{1}{4 e^2}F_{\mu\nu}(A)F^{\mu\nu}(A)-\frac{1}{4 e^2}F_{\mu\nu}(B)F^{\mu\nu}(B)\right]}\;\;,
\end{array}
\end{equation}
We now follow \cite{Mavromatos1} and replace the (2+2) degrees of freedom associated with the fields $A$ and $B$ with the (3+1) degrees of freedom associated with a massive bosonic field $B$ and a massless scalar field $\phi$. In this sense, the field $\phi$ can be thought as a Goldstone boson associated with the breaking of the $U(1)$ symmetry for the field $A$. Differently from a conventional BCS theory this happens without a local order parameter. The emergence of a mass for the boson field $A$ already is a signature of the physics associated with the Meissner effect. \\
We now introduce an antisymmetric tensor field $Z_{\mu\nu}$ through
\begin{equation}
\int \mathcal{D}Z\delta(Z_{\mu\nu}-F_{\mu\nu}(B))=1\;\;.
\end{equation}
The delta function can be represented as
\begin{equation}
\delta(T_{\mu\nu})=\frac{1}{2\pi}\int \mathcal{D}L_{\mu\nu}e^{i\int d^3 x L_\mu \epsilon^{\mu\alpha\beta}T_{\alpha\beta}}\;\;.
\end{equation}
By reabsorbing constant factors we get
\begin{equation}
\label{eq:middle}
\begin{array}{lll}
Z&=&\int \mathcal{D}A\mathcal{D}B\mathcal{D}Z\mathcal{D}L \text{exp}\left\{i\int d^3 x\left[ - \frac{n}{2\pi}\epsilon^{\lambda\mu\nu}A_\lambda Z_{\mu\nu}-\frac{1}{4 e^2}Z_{\mu\nu}Z^{\mu\nu}-\frac{1}{4 e^2}F_{\mu\nu}(A)F^{\mu\nu}(A)\right.\right.\\
&&\left.\left.+L_\mu \epsilon^{\mu\alpha\beta}(Z_{\alpha\beta}-F_{\mu\nu}(B))\right]\right\}\;\;.
\end{array}
\end{equation}
By using again the representation of the delta function given above, we now perform the integration over $B$ to
\begin{equation}
\begin{array}{lll}
Z&=&\int \mathcal{D}A\mathcal{D}Z\mathcal{D}L \delta(\epsilon_{\mu\alpha\beta}\partial^\alpha L^\beta)\text{exp}\left\{i\int d^3 x\left[ - \frac{n}{2\pi}\epsilon^{\lambda\mu\nu}A_\lambda Z_{\mu\nu}-\frac{1}{4 e^2}Z_{\mu\nu}Z^{\mu\nu}\right.\right.\\
&&\left.\left.-\frac{1}{4 e^2}F_{\mu\nu}(A)F^{\mu\nu}(A)+L_\mu \epsilon^{\mu\alpha\beta}Z_{\alpha\beta}\right]\right\}\;\;.
\end{array}
\end{equation}
The constraint $\epsilon_{\mu\alpha\beta}\partial^\alpha L^\beta=0$ can be implemented by imposing $L_\mu=\partial_\mu \phi$ where $\phi$ is a scalar field. In this way we get
\begin{equation}
Z=\int \mathcal{D}A\mathcal{D}Z\mathcal{D}\phi \text{exp}\left\{i\int d^3 x\left[ -\frac{1}{4 e^2}F_{\mu\nu}(A)F^{\mu\nu}(A)-\frac{1}{4 e^2}Z_{\mu\nu}Z^{\mu\nu}\right.\right.\left.\left.+(\partial_\mu\phi-\frac{n}{2\pi}A_\mu) \epsilon^{\mu\alpha\beta}Z_{\alpha\beta}\right]\right\}\;\;.
\end{equation}
The integration over $Z$ is a gaussian integral which leads to
\begin{equation}
\label{eq:phiTheory}
Z_\phi=\int \mathcal{D}A\mathcal{D}\phi e^{i\int d^3 x\left[ -\frac{1}{4 e^2}F_{\mu\nu}(A)F^{\mu\nu}(A)+2 e^2 (\partial_\mu\phi-\frac{n}{2\pi}A_\mu)^2 \right]}\;\;.
\end{equation}
From this we can see that, indeed, the field $A$ acquires a mass which breaks the gauge symmetry of the original model.\\
The charge and currents associated with the field are, by definition,
\begin{equation}
\label{eq:rhoJ}
\left\{\begin{array}{lll}
\rho&=&\frac{\delta \mathcal{L}_\phi}{\delta A_0}\\
\bf{J}_\text{em}&=&\frac{\delta \mathcal{L}_\phi}{\delta \bf{A}}\;\;,
\end{array}\right.
\end{equation}
%Now we have that
%\begin{equation}
%\rho=\frac{\delta \mathcal{L}}{\delta (\partial_0 \tilde{\phi})}\\
%\end{equation}
%where $\tilde{\phi}$ is such that $\partial_0\tilde{\phi}=\partial_0-\frac{n}{2\pi}A_0$ so that $\tilde{\phi}$ and $A_0$ are conjugate variables. The Hamilton equation of motion are
%\begin{equation}
%\partial_0\tilde{\phi}=\frac{\delta\mathcal{H}}{\delta \rho}
%\end{equation}
%and this gives
%\begin{equation}
%\label{eq:Vphi}
%V=\frac{\partial\tilde{\phi}}{\partial t}
%\end{equation}
%where $V$ is the voltage. So, the presence of steady state currents implies 
%\begin{equation}
%\label{eq:V0}
%V=0
%\end{equation}
%Another, maybe more precise, way to put it is to first 
where $\mathcal{L}_\phi$ is the Lagrangian associated with the partition function $Z_\phi$. We now observe that $\pi_\phi$, the momentum conjugate to the variable $\phi$ is
\begin{equation}
\begin{array}{lll}
\pi_\phi&=&\frac{\delta\mathcal{L}}{\delta\partial_0\phi}\\
&=&\frac{\delta\mathcal{L}}{\delta(\partial_0\phi-\frac{n}{2\pi}A_0)}\\
&=&\frac{2\pi}{n}\frac{\delta\mathcal{L}}{\delta(A_0)}\\
&=&\frac{2\pi}{n}\rho\;\;.
\end{array}
\end{equation}
This shows that the charge density $\rho$ is the canonical momentum conjugate to the field $\phi$.
The Hamilton equations of motion are $\partial_0\phi=\frac{\delta\mathcal{H}}{\delta \rho}$ and this gives
\begin{equation}
\label{eq:Vphi}
V=\frac{\partial\phi}{\partial t}\;\;,
\end{equation}
where $V$ is the voltage. So, the presence of steady state currents implies 
\begin{equation}
\label{eq:V0}
V=0\;\;.
\end{equation}
Since we have a situation with time independent currents and zero potential energy we can use the Drude formula
\begin{equation}
{\bf J}=\sigma{\bf{E}}\;\;,
\end{equation}
to describe this model with $\sigma=\infty$.
%\newpage
\subsection{Observables}
We now want to associate identities between fermionic physical observables to two key superconducting features: Meissner effect and infinite conductance.\\
To this end, let us add source terms $F_{\mu\nu}(A)\epsilon^{\mu\nu\lambda}J^A_\lambda$ and $F_{\mu\nu}(B)\epsilon^{\mu\nu\lambda}J^B_\lambda$ to the theory described by  Eq. (\ref{eq:bosonicTheory})
\begin{equation}
\begin{array}{lll}
Z_{\text{M}^2\text{BF}}(J^A,J^B)&=&\int \mathcal{D}A\mathcal{D}B \text{exp}\left\{i\int d^3 x\left[  \frac{n}{\pi}\epsilon^{\lambda\mu\nu}B_\lambda\partial_\mu A_\nu-\frac{1}{4 e^2}F_{\mu\nu}(A)F^{\mu\nu}(A)\right.\right.\\
&&\left.\left.-\frac{1}{4 e^2}F_{\mu\nu}(B)F^{\mu\nu}(B)+F_{\mu\nu}(A)\epsilon^{\mu\nu\lambda}J^A_\lambda+F_{\mu\nu}(B)\epsilon^{\mu\nu\lambda}J^B_\lambda\right]\right\}\;\;,
\end{array}
\end{equation}
These terms can be thought as a non-minimal coupling of a current $J$ to the field $A,B$. Alternatively, we can think of it as a minimal coupling $-2 A_\mu \tilde{J}^\mu$ between the field $A$ and a current $\tilde{J}=\epsilon_{\mu\nu\lambda}\partial^\nu J^\lambda$. \\
We now track the source term while following the steps that brought us from Eq. (\ref{eq:bosonicTheory}) to Eq. (\ref{eq:phiTheory}). In particular, this amounts in adding the term $F_{\mu\nu}(A)\epsilon^{\mu\nu\lambda}J^A_\lambda$ to each action and replacing, from Eq. (\ref{eq:middle}) on, the term $\epsilon^{\lambda\mu\nu}A_\lambda Z_{\mu\nu}\rightarrow \epsilon^{\lambda\mu\nu}(A_\lambda+J^B_\lambda) Z_{\mu\nu}$ which then leads to the generalization of  Eq. (\ref{eq:phiTheory})
\begin{equation}
\label{eq:ZphiJJ}
Z_\phi(J^A,J^B)=\int \mathcal{D}A\mathcal{D}\phi e^{i\int d^3 x\left[ -\frac{1}{4 e^2}F_{\mu\nu}(A)F^{\mu\nu}(A)+2 e^2 (\partial_\mu\phi-\frac{n}{2\pi}A_\mu+J^B_\mu)^2+F_{\mu\nu}(A)\epsilon^{\mu\nu\lambda}J^A_\lambda \right]}\;\;.
\end{equation}
Now, this is our final effective theory describing an effective electromagnetic potential inside our material. It has all the key features of superfluidity: the boson acquires a mass due to the interaction with the scalar field $\phi$ which acts analogously to a Goldstone boson. This effect describes the Meissner effect. On the other hand, in the steady state, we showed above that this system describes infinite conductivity compatible with the physics of a perfect conductor. 
The effective magnetic and electric fields inside the material are given by
\begin{equation}
\label{eq:ElMag}
\left\{\begin{array}{lll}
E^{i}&=&\frac{1}{2}\epsilon^{j\mu\nu}F(A)_{\mu\nu}\\
B_\text{mag}&=&\frac{1}{2}\epsilon^{0\mu\nu}F(A)_{\mu\nu}\;\;.
\end{array}\right.
\end{equation}
To establish a correspondence between these effects and fermionic observables, we want to find the fermionic version of both the electromagnetic charges and currents $(\rho, {\bf J}_\text{em})$ defined in Eq. (\ref{eq:rhoJ}) and the effective electromagnetic fields $(B_\text{mag},{\bf{E}})$ defined in Eq. (\ref{eq:ElMag}).\\
%The time dependence of the current $\bf{J}=\frac{\delta \mathcal{L}}{\delta \bf{A}}$ is constrained by Newton`s law:
%\begin{equation}
%{\bf{E}}=\Lambda\frac{d}{dt}\bf{J}
%\end{equation}
%We can compare this behaviour to the one the introduction of dissipation which is described by the Drude model as:
%\begin{equation}
%{\bf{E}}=\sigma\bf{J}
%\end{equation}
%On the other hand he magnetic field is expelled inside the material:
%\begin{equation}
%B=0
%\end{equation}
%We then want  relate the fields $(B,\bf{E})$ and the current $(\rho,\bf{J}_\text{em})$ to fermionic observables. 
From the definition in Eq. (\ref{eq:rhoJ}) and from the expression of the action associated with $Z_\phi$ in Eq. (\ref{eq:ZphiJJ}) we find  that
\begin{equation}
\left\{\begin{array}{lllll}
\rho&=&\frac{\delta \mathcal{L}}{\delta A_0}&=&\frac{2\pi}{n}\frac{\delta \mathcal{L}_\phi}{J_0^B}\\
J^i&=&\frac{\delta \mathcal{L}}{\delta A^k}&=&\frac{2\pi}{n}\frac{\delta \mathcal{L}_\phi}{\delta J_i^B}\;\;,
\end{array}\right.
\end{equation}
where $i=1,2$. 
This leads immediately to
\begin{equation}
\left\{\begin{array}{lll}
\langle\rho\rangle_\phi&=&\frac{2\pi}{n}\frac{\delta Z_\phi(J^A,J^B)}{\delta J_0^B}_{{\Big|}_{J^A,J^B=0}}\\
\langle J^i\rangle_\phi&=&\frac{2\pi}{n}\frac{\delta Z_\phi(J^A,J^B)}{\delta {\bf J}_i^B}_{{\Big|}_{J^A,J^B=0}}\;\;.
\end{array}\right.
\end{equation}
But, from the duality relation $Z_\phi=Z_{\text{M}^2BF}$ proved above we also have
\begin{equation}
\left\{\begin{array}{lll}
\langle\rho\rangle_\phi&=&\frac{2\pi}{n}\frac{\delta Z_{\text{M}^2BF}(J^A,J^B)}{\delta J_0^B}_{{\Big|}_{J^A,J^B=0}}=\langle\epsilon^{\mu\nu 0}F_{\mu\nu}(B)\rangle_{\text{M}^2BF}\\
\langle J^i\rangle_\phi&=&\frac{2\pi}{n}\frac{\delta Z_{\text{M}^2BF}(J^A,J^B)}{\delta J_i^B}_{{\Big|}_{J^A,J^B=0}}=\langle\epsilon^{\mu\nu i}F_{\mu\nu}(B)\rangle_{\text{M}^2BF}\;\;.
\end{array}\right.
\end{equation}
This relation, together with the equivalence between (double-Maxwell)-BF and (double CP)-BF theories and with the fermionization rules given in Eq. (\ref{eq:fermionizationRules1}) implies
\begin{equation}
\left\{\begin{array}{lll}
\langle\rho\rangle_\phi&=&\frac{\sqrt{2} \pi}{n}\sum_j \langle J^j_0\rangle_{\text{F}}\\
\langle J_\text{em}^i\rangle_\phi&=&\frac{\sqrt{2} \pi}{n}\sum_j\langle J^j_i\rangle_{\text{F}}\;\;,
\end{array}\right.
\end{equation}
where, we remark that, in the symbol $J^j_i$, the label $j$ corresponds to the layer index, while $i$ corresponds to a spatial component ($i=1,2$) of the current.\\ 
On the other hand, the expectation values for the effective electric and magnetic fields can be  simply written as
\begin{equation}
\left\{\begin{array}{lll}
\langle E^i\rangle&=&\frac{1}{2}\frac{\delta Z_\phi (J^A,J^B)}{\delta J^A_i}_{{\Big|}_{J^A,J^B=0}}\\
\langle B_\text{mag}\rangle&=&\frac{1}{2}\frac{\delta Z_\phi (J^A,J^B)}{\delta J^A_0}_{{\Big|}_{J^A,J^B=0}}\;\;,
\end{array}\right.
\end{equation}
which using the duality, leads to
\begin{equation}
\begin{array}{lllll}
\langle E^i\rangle&=&\frac{1}{2}\frac{\delta Z_{\text{M}^2BF}(J^A,J^B)}{\delta J^A_i}_{{\Big|}_{J^A,J^B=0}}&=&\langle\epsilon^{\mu\nu i}F_{\mu\nu}(A)\rangle_{\text{M}^2BF}\\
\langle B_\text{mag}\rangle&=&\frac{1}{2}\frac{\delta Z_{\text{M}^2BF}(J^A,J^B)}{\delta J^A_0}_{{\Big|}_{J^A,J^B=0}}&=&=\langle\epsilon^{\mu\nu 0}F_{\mu\nu}(A)\rangle_{\text{M}^2BF}\;\;.
\end{array}
\end{equation}
Now, thanks to the fermionization rules, we have
\begin{equation}
\left\{\begin{array}{lll}
\langle B_\text{mag}\rangle_\phi&=&-\frac{\sqrt{2} \pi}{n}\sum_j \langle J^{5 j}_0\rangle_{\text{F}}\\
\langle E^i\rangle_\phi&=&-\frac{\sqrt{2} \pi}{n}\sum_j\langle J^{5 j}_i\rangle_{\text{F}}\;\;.
\end{array}\right.
\end{equation}
All these findings can be nicely summarized in the following table.
\begin{center}
\begin{tabular}{|cc|}
\hline
Electromagnetic Quantities & Fermionic Oservables\\
\hline
$\langle\rho\rangle_\phi$&$\sum_j\langle J^j_0\rangle_F$\\
$\langle{\bf{J}}_\text{em}\rangle_\phi$&$\sum_j\langle{\bf{J}}^j\rangle_F$\\
$\langle B_\text{mag}\rangle_\phi$&$\sum_j\langle J^{5 j}_0\rangle_F$\\
$\langle{\bf E}\rangle_\phi$&$\sum_j\langle{\bf J}^{5 j}\rangle_F$\\
\hline
\end{tabular}
\end{center}
We now use these correspondences to map the superfluidity effects to identities among fermionic observables.\\
The (effective) Meissner effect is characterized by the expulsion of the (effective) magnetic field from the sample and, in our case, it reads
\begin{equation}
B_\text{mag}=0\;\;,
\end{equation}
within a penetration depth from the boundary given by $\lambda\propto \frac{1}{e^2}$ \cite{Mavromatos1}, which, by using Eq. (\ref{eq:ee}) can be written in terms of the interaction strength $g$ as $\lambda\propto g^2$.
From the table above we see that this implies
\begin{equation}
\label{eq:MessnierA}
\sum_j\langle J^{5 j}_0\rangle_F=0\;\;.
\end{equation}
which is an identity that has to be satisfied in the original fermionic model.\\
On the other hand, an infinite conductivity is represented by a Drude formula ${\bf J}=\sigma {\bf E}$ with $\sigma=\infty$ and its correspondent fermionic identity is given by
\begin{equation}
\label{eq:tricky}
\sum_j \langle{\bf J}^j\rangle_F=\sigma\sum_j \langle{\bf J}^{5 j}\rangle_F\;\;.
\end{equation}
We note that superconducting currents only flow within a distance $\lambda\propto g^2$ from the boundary of the sample. This means that the superfluidity of our model has a tunable penetration depth (depending on the interaction strength $g$). On can use this feature to insure that the identity in Eq. (\ref{eq:tricky}) is valid inside the bulk of the material where the fermionization rules hold.\\
%In summary
%\begin{center}
%\begin{tabular}{|ccc|}
%\hline
%Physical Effect&Relation&Fermionic Relation\\
%\hline
%No dissipation&${\bf E}=\Lambda\frac{d}{dt}{\bf J}$&$<{\bf J}^5>_F=\Lambda \frac{d}{dt}<{\bf J}>_F$\\
%Dissipation&${\bf J}=\sigma {\bf E}$&$<{\bf J}>_F=\sigma<{\bf J}^5>_F$\\
%Messnier Effect&$B=0$&$<J^5_0>_F=0$\\
%\hline
%\end{tabular}
%\end{center}
The observable identity in Eq. (\ref{eq:MessnierA}) is consistent with the skyrmionic interpretation of the effective superfludity proposed in this article. In fact, the skyrmion currents for the theory in Eq. (\ref{eq:2CPCS}) can be written as \cite{Fradkin1}
\begin{equation}
J^{\mu \pm}_{S}=\frac{1}{2\pi}\epsilon^{\mu\nu\lambda}\partial_\nu A_\lambda^{\pm}\;\;.
\end{equation}
 Now, by using the fermionization rules and the change of variables in Eq.(\ref{eq:changeOfVariablesCPCS}) we have
\begin{equation}
\label{eq:skyrmionsFermions}
\left\{\begin{array}{lll}
\langle J^{\mu +}_S\rangle_{(\text{CP-CS})^2}&=&\frac{1}{n\sqrt{2}}\sum_j(\langle J_j^\mu\rangle_F-\langle J_j^{\mu 5}\rangle_F)\\
\langle J^{\mu -}_S\rangle_{(\text{CP-CS})^2}&=&-\frac{1}{n\sqrt{2}}\sum_j(\langle J_j^\mu\rangle_F+\langle J_j^{\mu 5}\rangle_F)\;\;,
\end{array}\right.
\end{equation}
which connects the skyrmionic currents (which involves bosons for $n$ even) to the fermionic observables.\\
Now, from Eq. (\ref{eq:O3Hopf}) we can see that the two skyrmions have opposite topological charge $Q_T$
\begin{equation}
\label{eq:QTplusQTminus}
Q_T^+=-Q_T^-\;\;,
\end{equation}
where
\begin{equation}
\label{eq:QQ}
Q^\pm_T=\int d^2 x J^{\pm0}_S\;\;.
\end{equation}
But, this gives us another way to prove Eq. (\ref{eq:MessnierA}), namely
\begin{equation}
\begin{array}{lll}
0&\eqtext{\text{Eq. }\ref{eq:QTplusQTminus}}&-\frac{n}{\sqrt{2}}(Q_T^++Q_T^-)\\
&\eqtext{\text{Eq. } \ref{eq:QQ}}&-\frac{n}{\sqrt{2}}\int d^2 x\left(\langle J^{+0}_S\rangle_{(\text{CP-CS})^2}+\langle J^{-0}_S\rangle_{(\text{CP-CS})^2}\right)\\
&\eqtext{\text{Eq. } \ref{eq:skyrmionsFermions}}&\int d^2 x \sum_j \langle J^{5 j}_0\rangle_F\;\;,
\end{array}
\end{equation}
which is consistent with Eq. (\ref{eq:Messnier}) which was derived a consequence of the effective Meissner effect.
\section{Bosonization}
We now want to obtain the mapping between the chiral-invariant Thirring model and the (double Maxwell)-BF theory through a generalization of the bosonization techniques introduced in \cite{Fradkin2}. 
Following what done before, we stress that we are going to treat each Fermi point independently. We then omit the momentum space label $\pm$ with the agreement that all the following is valid around each Fermi point.\\
Our starting point is the fermionic model given in Eq. (\ref{eq:S}) which we rewrite here
\begin{equation}
S=S_0+S_I\;\;,
\end{equation}
with
\begin{equation}
\begin{array}{lll}
S_0&=&\int d^3 x\bar{\Psi}_1(i c\slashed{\partial}-mc^2)\Psi_1+\bar{\Psi}_2(i c\slashed{\partial}-mc^2)\Psi_2\\
S_I&=&\int d^3 x\frac{g^2}{2}[(\sum_j J^j_\mu)^2+(\sum_j J^{j5}_\mu)^2]\;\;.
\end{array}
\end{equation}
We now use the Hubbard-Stratonovich transformation to write
\begin{equation}
\begin{array}{lll}
e^{iS_I}&=&\exp \frac{i g^2}{2}[(\sum_j J^j_\mu)^2+(\sum_j J^{j5}_\mu)^2]\\
&=&\int \mathcal{D}a \mathcal{D}b \;e^{i\int d^3 x (\frac{1}{2g^2}a_\mu a^\mu+ \sum_j J^j_\mu a^\mu+\frac{1}{2 g^2}b_\mu b^\mu+ \sum_j J^{j5}_\mu b^\mu)}\;\;.
\end{array}
\end{equation}
The full action of the model can then be written as
\begin{equation}
\begin{array}{lll}
S&=&\int d^3 x\left[\Psi_1^\dagger(i c\slashed{\partial}-mc^2)\Psi_1+\Psi_2^\dagger(i c\slashed{\partial}-mc^2)\Psi_2+\frac{1}{2 g^2} a_\mu a^\mu+\frac{1}{2 g^2}b_\mu b^\mu\right.\\
&&\left.+ J^{(1)}_\mu a^\mu+J^{(1)5}_\mu b^\mu++J^{(2)}_\mu a^\mu+ J^{(2)5}_\mu b^\mu\right]\;\;,
\end{array}
\end{equation}
where $J^{(1,2)}=\Psi_{1,2}^\dagger\gamma^\mu\Psi_{1,2}$ and $J^{(1,2)5}=\Psi_{1,2}^\dagger\gamma^5\gamma^\mu\Psi_{1,2}$. By choosing $c=1$, we can write the partition function of the model as
\begin{equation}
Z_F=\int \mathcal{D}a\mathcal{D}b\; e^{iS}\;\;,
\end{equation}
where
\begin{equation}
\begin{array}{lll}
S&=&S_\Psi^1+S_\Psi^2+S_F\;\;,
\end{array}
\end{equation}
with
\begin{equation}
\begin{array}{lll}
S^1_\Psi&=&\int d^3 x\left[\Psi_1^\dagger(i\slashed{\partial}+\slashed{a}+\gamma^5\slashed{b}-m)\Psi_1\right]\\
&&\\
S^2_\Psi&=&\int d^3 x\left[\Psi_2^\dagger(i\slashed{\partial}+\slashed{a}+\gamma^5\slashed{b}-m)\Psi_2\right]\\
S_F&=&\int d^3 x\left[\frac{1}{ 2 g^2}a_\mu a^\mu+\frac{1}{ 2 g^2}b_\mu b^\mu\right]\;\;.
\end{array}
\end{equation}
Dimensional analysis shows that $[\Psi]=L^{-1}$, $[m]=L^{-1}$, $[g^2]=L$, $[a]=[b]=L^{-1}$.
To begin we can integrate out the fermionic degrees of freedom to get
\begin{equation}
\begin{array}{lll}
\int \mathcal{D}\Psi_1 \mathcal{D}\Psi_1^\dagger \mathcal{D}\Psi_2 \mathcal{D}\Psi_2^\dagger\;e^{i(S^1_\Psi+S^2_\Psi)}=e^{iS_\text{eff}}\;\;,
\end{array}
\end{equation}
where
\begin{equation}
\begin{array}{lll}
S_\text{eff}=-i n\log{\text{det}(\slashed{\partial}+\slashed{a}+\gamma^5\slashed{b}-m)}=\frac{n}{2\pi}\frac{m}{|m|}\epsilon^{\mu\nu\lambda}\int b_{\mu}\partial_{\nu} a_{\lambda}+O(\frac{\partial}{m c})\;\;.
\end{array}
\end{equation}
%where, importantly, $n=2$ is the number of spinors in our model.
We can now rewrite the low-energy limit of $S$ as
\begin{equation}
S=\int d^3 x \left[\frac{n}{2\pi}\epsilon^{\lambda\mu\nu}b_\lambda\partial_\mu a_\nu+\frac{1}{2 g^2}a_\mu a^\mu+\frac{1}{2 g^2}b_\mu b^\mu\right]\;\;.
\end{equation}
By changing variables as $a_\mu=\frac{1}{2}(t_\mu+s_\mu)$ and $b_\mu=\frac{1}{2}(t_\mu-s_\mu)$ we get
\begin{equation}
S=\int d^3 x \left[\frac{n}{8\pi}\epsilon^{\lambda\mu\nu}t_\lambda\partial_\mu t_\nu-\frac{n}{8\pi} \epsilon^{\lambda\mu\nu}s_\lambda\partial_\mu s_\nu+\frac{1}{4g^2}s_\mu s^\mu+\frac{1}{4g^2}t_\mu t^\mu\right]\;\;.
\end{equation}
We now define the following interpolating action
\begin{equation}
\begin{array}{lll}
S_I&=&\int d^3 x (\frac{1}{4g^2}s_\mu s^\mu+q_1 \epsilon^{\lambda\mu\nu}s_\lambda\partial_\mu A^+_\nu-q_2 \epsilon^{\lambda\mu\nu}A^+_\lambda\partial_\mu A_\nu^+)\\
&&+\int d^3 x (\frac{1}{4g^2}t_\mu t^\mu+q_1\epsilon^{\lambda\mu\nu}t_\lambda\partial_\mu A^-_\nu+q_2\epsilon^{\lambda\mu\nu}A^-_\lambda\partial_\mu A_\nu^-)\;,
\end{array}
\end{equation}
having
\begin{equation}
\label{eq:D1}
\int \mathcal{D}A^+\mathcal{D}A^- e^{i S_I}=e^{i(\frac{1}{4g^2}s_\mu s^\mu+\frac{1}{4g^2}t_\mu t^\mu- \frac{q^2_1}{4 q_2}\epsilon^{\lambda\mu\nu}s_\lambda\partial_\mu s_\nu+ \frac{q^2_1}{4 q_2}\epsilon^{\lambda\mu\nu}t_\lambda\partial_\mu t_\nu)}\;\;,
\end{equation}
which is our original theory for $\frac{q^2_1}{4 q_2}=\frac{n}{8 \pi}$ while
\begin{equation}
\label{eq:D2}
\int \mathcal{D}s\mathcal{D}t\; e^{i S_I}=e^{i \int d^3x(q_2 A^+_\lambda\partial_\mu A_\nu^+ -q_2 A^-_\lambda\partial_\mu A_\nu^-+(2 q_1 g^2)^2(F^+_{\mu\nu}F^{+\mu\nu})+(2 q_1 g^2)^2(F^-_{\mu\nu}F^{-\mu\nu}))}\;\;.
\end{equation}
By defining new fields $A$ and $B$ as $A^+_\mu=A^\mu+B^\mu$ and $A^-_\mu=A^\mu-B^\mu$, we get the dual theory with action
\begin{equation}
S=\int d^3 x\left[\frac{n}{\pi}\epsilon^{\lambda\mu\nu}B_\lambda\partial_\mu A_\nu+\frac{1}{4 e^2}F(A)_{\mu\nu}F(A)^{\mu\nu}+\frac{1}{4 e^2}F(B)_{\mu\nu}F(B)^{\mu\nu}\right]\;\;,
\end{equation}
(where $e^{2}=\frac{2 \pi q_2}{4 n q_1 g^2}=\frac{\pi^{2}q_{1}}{n^{2}g^{2}}$),
which is nothing but the (double Maxwell)-BF theory.  Then we have found the wanted low-energy correspondence between different theories in the low-energy limit by using functional bosonization
\begin{equation}
\begin{array}{lll}
Z_{M^2-BF}\approx Z_F\;.
\end{array}
\end{equation}


\begin{thebibliography}{25}

\bibitem{Ando}
Y. Ando, J. Phys. Soc. Jpn. \textbf{82}, 102001 (2013).

\bibitem{Qi1}
X.-L. Qi and S.-C. Zhang, Rev. Mod. Phys. \textbf{83}, 1057 (2011).

\bibitem{Palumbo1}
G. Palumbo and J. K. Pachos, Phys. Rev. Lett. \textbf{110}, 211603 (2013).


\bibitem{Palumbo2}
G. Palumbo and J. K. Pachos, Phys. Rev. D \textbf{90}, 027703(R) (2014).

\bibitem{Cirio}
M. Cirio, G. Palumbo and J. K. Pachos, Phys. Rev. B \textbf{90}, 085114 (2014).

\bibitem{Fradkin1}
E. Fradkin, {\em Field Theories of Condensed Matter Physics}, Cambridge University Press (2013).

\bibitem{Moore1}
G. Y. Cho and J. E. Moore, Ann. of Phys. \textbf{326}, 1515 (2011).


\bibitem{Hansson}
T. H. Hansson, V. Oganesyan, and S. L. Sondhi, Ann. of Phys. \textbf{313}, 497 (2004).

\bibitem{Haldane1}
F. D. M. Haldane, Phys. Rev. Lett. \textbf{50}, 1153 (1983).

\bibitem{Wen1}
X. Chen, Z.-C. Gu, Z.-X. Liu and X.-G. Wen, Phys. Rev. B \textbf{87}, 155114 (2013).

\bibitem{Xu}
C. Xu, Phys. Rev. B \textbf{87}, 144421 (2013).


\bibitem{Abanov1}
A. G. Abanov and P. B. Wiegmann, Nucl. Phys. B \textbf{570}, 685 (2000).


\bibitem{Wilczek}
F. Wilczek and A. Zee, Phys. Rev. Lett. \textbf{51}, 2250 (1983).

\bibitem{Polyakov}
A.M. Polyakov, Mod. Phys. Lett. A \textbf{3}, 325 (1988).


\bibitem{Abanov2}
A. G. Abanov and P. B. Wiegmann, Phys. Rev. Lett. \textbf{86}, 1319 (2001).


\bibitem{Senthil1}
T. Grover and T. Senthil, Phys. Rev. Lett. \textbf{100}, 156804 (2008).

\bibitem{Baskaran}
G. Baskaran, arXiv:1108.3562.


\bibitem{Mavromatos1}
N. Dorey and N. E. Mavromatos, Nucl. Phys. B \textbf{386}, 614 (1992).


\bibitem{Semenoff}
G. W. Semenoff and N. Weiss, Phys. Lett. B \textbf{250}, 117 (1990).

\bibitem{Mavromatos2}
N. E. Mavromatos and M. Ruiz-Altaba, Phys. Lett. A \textbf{142}, 419 (1989).


\bibitem{Eguchi}
T. Eguchi and H. Sugawara, Phys. Rev. D \textbf{10}, 4257 (1974).

\bibitem{Huerta}
L. Huerta and M. Ruiz-Altaba, Phys. Lett. B \textbf{216}, 371 (1989).

\bibitem{Fradkin2}
E. Fradkin, and F. A. Schaposnik, Phys. Lett. B \textbf{338}, 253 (1994).


\bibitem{Haldane2}
F. D. M. Haldane, Phys. Rev. Lett. \textbf{61}, 2015 (1988).

\bibitem{SM}
See Supplemental Material.

\bibitem{Karabali}
M. J. Bowick, D. Karabali and L. C. R. Wijewardhana, Nucl. Phys. B \textbf{271}, 417 (1986).


\bibitem{Schakel}
A. M. J. Schakel, Phys. Rev. D \textbf{44}, 1198 (1991).

\bibitem{Sim1}
I. Georgescu, S. Ashhab and F. Nori, Rev. Mod. Phys. \textbf{86}, 153 (2014).

\bibitem{Sim2}
I. Buluta and F. Nori, Science \textbf{326}, 108-111 (2009).

\bibitem{Sim3}
J.Q. You, Z.D. Wang, W. Zhang and F. Nori, Scientific Reports \textbf{4}, 5535 (2014).

\bibitem{Read}
N. Read and D. Green, Phys. Rev. B \textbf{61}, 10267 (2000).

\bibitem{Nori}
A. L. Rakhmanov, A. V. Rozhkov and F. Nori, Phys. Rev. B \textbf{84}, 075141 (2011).

\bibitem{Nori2}
R.S. Akzyanov, A.V. Rozhkov, A.L. Rakhmanov and  F. Nori, Phys. Rev. B \textbf{89}, 085409 (2014).

\end{thebibliography}
\end{document}